\documentclass[a4paper,12pt]{article}
\usepackage{amssymb,amsmath}
\usepackage{graphics}
\def\beq{\begin{equation}}
\def\eeq{\end{equation}}
\def\bea{\begin{eqnarray}}
\def\eea{\end{eqnarray}}
\def\nn{\nonumber}

\def\s#1{\bigg\lgroup #1 \bigg\rgroup}
\def\a{\alpha({\tilde{\beta}})}
\def\ag{\alpha({\hat{\beta}})}

\def\u{\mathcal{L}}
\def\zn{N_{BG}({\beta})}
\def\ab{({\tilde{\beta}}\mu)}
\def\up{\Upsilon_{BG}({\hat{\beta}})}
\def\upn{\Upsilon_{BG}(\beta)}
\def\ab{({\tilde{\beta}}\mu)}
\def\hb{({\hat{\beta}}\mu)}
\def\bmu{\beta\mu}
\def\b{{\alpha(\beta)}}
\def\bt{\beta}
\def\bb{(\beta \mu)}
\def\tb{{\hat{\beta}}}
\def\hb{{\hat{\beta}}}
\def\omt{\omega{(\hb)}}
\def\om{\omega{(\beta)}}
\makeatletter
  \def\@cite#1#2{${\mbox{#1\if@tempswa , #2\fi}}$}
\makeatother

\makeatletter
  \def\@biblabel#1{$^{\mbox{#1}}$}
\makeatother
\begin{document}
%
%
%
%
\thispagestyle{empty}

\vspace*{3cm}
\begin{center}
{\LARGE\sf A class of energy based ensembles in Tsallis statistics} \\

\bigskip\bigskip
R. Chandrashekar${}^{1}$ and S.S. Naina Mohammed${}^{2}$\\

\bigskip
\textit{
${}^{1}$Department of Theoretical Physics, \\
University of Madras, \\
Guindy Campus, Chennai 600 025, India \\}
\bigskip
\textit{
${}^{2}$Department of Education in Science and Mathematics\\
Regional Institute of Education\\
Mysore 570 006, India.
}
\end{center}

\vfill
\begin{abstract}
A comprehensive investigation is carried out on the class of energy based ensembles. The eight ensembles are divided into two main classes. In the isothermal class of ensembles the individual members are at the same temperature.  A unified framework is
evolved to describe the four isothermal ensembles. Such a description is provided both in the second and the third constraint 
formalisms. The isothermal-isobaric, grandcanonical and the generalized ensembles are illustrated
through a study of the classical nonrelativistic and the extreme relativistic ideal gas models. In the adiabatic class of ensembles the individual members of the ensemble have the same value of the heat function and a unified formulation to described all the four
ensembles is given. The nonrelativistic and the extreme relativistic ideal gases are studied in the isoenthalpic-isobaric ensemble, the adiabatic ensemble with number fluctuations, and, the adiabatic ensemble with number and particle fluctuations.
\end{abstract}
PACS Number(s): 05.20.-y, 05.70 
\newpage
\setcounter{page}{1}
%
%
%
\setcounter{equation}{0}
\section{Introduction}
\label{Intro}
Boltzmann-Gibbs extensive statistical mechanics was generalized
by Tsallis through the use of a deformed logarithm function in the
entropic expression [\cite{CT1}].  The functional form of the entropy defined in
terms of the deformed logarithm is
\beq
S_{q} = -k \sum_{i} {\mathfrak{p}}_{i}^{q} \ln_{q} {\mathfrak{p}}_{i},
\qquad
q \in \mathbb{R}_{+},
\label{q_entr}
\eeq
where $k$ is the Boltzmann constant and ${\mathfrak{p}}_{i}$ is the
probability associated with the $i$-th microstate.
The positive real condition imposed on the nonextensivity parameter $q$
ensures the stability of the Tsallis entropy [\cite{SA}]. The $q$-deformed logarithm
made use of in (\ref{q_entr}) and its inverse the $q$-exponential are defined
below
\beq
\ln_{q}x = \frac{x^{1-q}-1}{1-q},\qquad
\exp_{q}(x) = [1+(1-q)x]^{\frac{1}{1-q}}_{+},
\eeq
where we have used the notation $[x]_{+} =$ max $\{0,x\}$. In the
$q \rightarrow 1$ limit, (\ref{q_entr}) reduces to the extensive Boltzmann-Gibbs
entropy. A detailed account on the properties of the $q$-entropy and its
applications may be found in [\cite{T2009}].

\par

To describe a system in thermodynamic equilibrium with its surroundings,
we need three macroscopic variables corresponding to thermal, mechanical and
chemical equilibrium.  A system with fixed values of these parameters can exist in
different microstates. A collection of systems existing in all the possible microstates
corresponding to the same value of the macroscopic variables is called an ensemble.
Each of these macroscopic parameters can be chosen between an extensive and an
intensive variable. Thus we have eight different ways of choosing the three variables
and consequently eight different ensembles. The eight ensembles can be divided into two
different classes namely the isothermal class and the adiabatic class. In the isothermal
class of ensembles, the thermal equilibrium variable is the temperature and the
individual members of the ensemble are at the same temperature. The canonical, isothermal-isobaric
[\cite{EG}-\cite{WBB}], grandcanonical and the generalized ensemble [\cite{EG},\cite{H56}] belong to the
isothermal class. Each one of these isothermal ensembles is characterized by an heat function and a free energy.
A general expression for the heat function and the generalized free energy are given below.
\bea
\mathfrak{H} &=& U + \sum_{\{\ell\}} x_{\ell} X_{\ell}
\label{hf_def} \\
\mathcal{F} &=& U - T S + \sum_{\{\ell\}} x_{\ell} X_{\ell}
\label{gfe_def}
\eea
where $X$ is an extensive thermodynamic quantity whose intensive conjugate is $x$.
The index $\ell$ can take three different values of either $0$, $1$ or $2$.  In the $\ell=0$
case the internal energy and the Helmholtz free energy are the heat function and the free energy
respectively. The ensemble pertaining to this situation is the canonical ensemble ($N$,$V$,$T$).
Both the isothermal-isobaric ($N$,$P$,$T$) and the grandcanonical ($\mu$,$V$,$T$) ensemble relate to
the $\ell=1$ case in (\ref{hf_def}) and (\ref{gfe_def}).  The enthalpy $H = U + PV$ and the
Gibbs free energy $G = U - TS + PV$ are the heat function and the free energy corresponding to the
isothermal-isobaric ensemble. For the grandcanonical ensemble the Hill energy $\mathsf{L} = U - \mu N$
and the $\Phi = U - TS - \mu N$ are the heat function and the free energy respectively.  The $\ell=2$
case describes the completely open system ($\mu$,$P$,$T$) and the ensemble is usually referred to as the
generalized ensemble. The $\mathsf{R}$ function $\mathsf{R} = U + PV - \mu N$ and
$\mathcal{E} = U + PV - TS - \mu N$ are the heat function and the free energy of the generalized ensemble.
The adiabatic class of ensembles arise when the thermal equilibrium happens with respect to the heat function
which is an extensive quantity. Based on the four different heat functions namely $U$,$H$,$\mathsf{L}$ and
$\mathsf{R}$ we have four different ensembles. The well known microcanonical ($N$,$V$,$U$) ensemble.
The isoenthalpic-isobaric ($N$,$P$,$H$) ensemble introduced in [\cite{JH1}] and later extensively studied in
references [\cite{JR1},\cite{JR3}]. More recently the adiabatic ensemble with number fluctuations ($\mu$,$V$,$\mathsf{L}$)
and the adiabatic ensemble with number and volume fluctuations ($\mu$,$P$,$\mathsf{R}$) were introduced through
references [\cite{JR2}] and [\cite{JR4}] respectively. All the four adiabatic ensembles were treated in a unified way in
[\cite{HG1}].  Of the eight ensembles, the microcanonical, the canonical and the grandcanonical ensembles have been
extensively studied and are well known in statistical mechanics. But the other five ensembles are also of fundamental
importance in studying physical problems whose experimental conditions emulate the conditions of these ensembles. For example the isothermal-isobaric and the isoenthalpic-isobaric ensembles are used to study fluids which are confined at constant pressure.  A
molecular dynamic simulation for these ensembles was evolved in [\cite{HA}]. The generalized ensemble has been
studied in [\cite{H63}]. Applications of the generalized ensemble include lattice gas models,
and a system of polymer molecule of $N$ monomers held together by weak forces such that the number $N$
fluctuates [\cite{H63}].

\par

An investigation on the ensemble formulation of nonextensive statistical mechanics based on
$q$-entropy (\ref{q_entr}) was initiated in [\cite{CT1}] through the study of two level
system in the microcanonical and the canonical ensembles. A further detailed study of the
canonical ensemble was carried out in references [\cite{CUR},\cite{TMP}] wherein
the second and the third constraint models were introduced. In the second constraint method
[\cite{CUR}] we use the unnormalized $q$-expectation value which, for a given observable $O$ reads:
\beq
\langle O\rangle_{q}^{(2)} = \sum_{j}\Big(
                             {\mathfrak{p}}_{j}^{(2)}(\beta)\Big)^{q} \,O_{j},
\qquad
\langle 1\rangle_{q}^{(2)} \equiv {\mathfrak{c}}^{(2)}(\beta)
                           = \sum_{j}\Big(
                             {\mathfrak{p}}_{j}^{(2)}(\beta)\Big)^{q},
\label{q_O_2}
\eeq
where ${\mathfrak{p}}_{j}^{(2)}$ is the ensemble probability of the microstate $j$ in
the second constraint. Though the thermodynamic Legendre structure is preserved in
the second constraint, the formalism suffered from certain disadvantages.  The unit
operator does not preserve its norm for an arbitrary value of $q$ and, the form of the
energy conservation principle in the macroscopic and microscopic limits varied.
In order to overcome the disadvantages in the second constraint, an appropriately
normalized form of the $q$-expectation value known as the third constraint method was
introduced in [\cite{TMP}]. The definition of the $q$-expectation in the third constraint formalism
is
\beq
\langle O\rangle_{q}^{(3)} = \frac{\sum_{j}
                             \Big({\mathfrak{p}}_{j}^{(3)}(\beta)\Big)^{q}
                             \,O_{j}}{{\mathfrak{c}}^{(3)}(\beta)},
\qquad
{\mathfrak{c}}^{(3)}(\beta) = \sum_{j} \Big({\mathfrak{p}}_{j}^{(3)}
                              (\beta)\Big)^{q}.
\label{q_O_3}
\eeq
Though the third constraint method cleared all the problems previously encountered in
the second constraint method, the probabilities became implicit quantities rendering the
calculations difficult. But fortunately, the second and the third constraint models
can be interrelated via a temperature transformation relation which helps us to override the
difficulties caused by the implicitness in the probabilities and the thermodynamic variables.
The various applications of the canonical formulation may be found in [\cite{SA1},\cite{CCN1}]. A similar
formulation of the grandcanonical ensemble was first done in the second constraint [\cite{C1995}] and
later in the third constraint [\cite{JN}]. A detailed investigation of the microcanonical ensemble based
on the $q$-entropy (\ref{q_entr}) was carried out in [\cite{CVG}] and the classical ideal gas was
discussed as an application. Thus in the nonextensive statistical mechanics, based on
$q$-entropy only three ensembles, the canonical [\cite{CT1},\cite{CUR},\cite{TMP}]
the microcanonical [\cite{CT1},\cite{CVG}] and the grandcanonical [\cite{C1995},\cite{JN}] have
been investigated so far.

\par

In the current work we present two unified frameworks one for the isothermal class of ensembles and the
other for the adiabatic class of ensembles. The unified framework pertaining to the isothermal class of ensembles
has been presented for both the second and the third constraint formalisms though the latter is the currently
accepted formulation. Adequate explanation for providing a unified formulation for the second constraint formalism
is discussed at the relevant places in the article. A generalized formulation of the temperature transformation relation has been
constructed to interrelate the second and the third constraint formulations.
An exact solution of the specific heats at constant pressure and at constant volume corresponding to
both the classical nonrelativistic ideal gas and the extreme relativistic ideal gas has been obtained in
the isothermal isobaric ensemble. Contrarily obtaining an exact solution to these ideal gas models in the
grandcanonical and the generalized ensemble appears to be difficult.  We make use of the technique of
disentangling of $q$-exponential into a series of ordinary exponentials [\cite{CQ}] to construct a perturbative
approach to the problem.  Such an approach has already been used in the context of the canonical ensembles
 [\cite{CCN1},\cite{CCN2}].  In the case of grandcanonical ensemble, the perturbative terms up to second order
in the expansion parameter $(1-q)$ has been retained. The perturbative terms have been displayed
up to first order in the expansion parameter $(1-q)$ in the case of the generalized ensemble.
The limitations imposed on the order of $(1-q)$ is only for the sake of simplicity and in practice
the method could be extended to any arbitrary order in the perturbative parameter. A unified framework
to describe all the four adiabatic ensembles in the nonextensive $q$-statistics has been put forward.
A $q$-generalization of the equipartition theorem and the virial theorem have also been discussed. The specific
heat at constant pressure and at constant volume corresponding to the nonrelativistic and extreme relativistic
models of ideal gas have been exactly computed. Since a summation over the number of particles could not
be evaluated in the ($\mu$,$V$,$\mathsf{L}$) and the ($\mu$,$P$,$\mathsf{R}$) adiabatic ensembles,
the phase space volume and the equation of state are expressed as formal sums. Throughout the article we
identify the temperature with the Lagrange multiplier corresponding to the constraint on the internal energy.
This is because if we assume the notion of quasireversibility and simultaneously implement the first
law of thermodynamics the Tsallis entropy becomes identical to the thermodynamic Clausius entropy [\cite{A06}]
only when the Lagrange multiplier $\beta$ is associated with the inverse temperature.

\par

The plan of this article is as follows:
The unified framework for the isothermal ensembles is presented in Section \ref{Isothermal}, wherein the second
and third constraint formulations are described as separate subsections.  Using the classical nonrelativistic and extreme
relativistic ideal gases the isothermal-isobaric ensemble, the grandcanonical ensemble and the
generalized ensemble are illustrated in Sections \ref{isobaric}, \ref{GCE} and \ref{GE}.  A generalized
formulation of the adiabatic ensembles is given in Section \ref{Adiabatic}.  The ideal gas model in the
isoenthalpic-isobaric ensemble have been worked out in Section \ref{NPH}.  In Sections \ref{LVmu} and
\ref{RPmu} the ($\mu$,$V$,$\mathsf{L}$) and the ($\mu$,$P$,$\mathsf{R}$) ensembles are discussed.
We present our concluding remarks in Section \ref{remarks}.
%
%
%
%
\setcounter{equation}{0}
\section{Isothermal ensemble}
\label{Isothermal}
The isothermal class of ensembles is used to  study systems in which the thermal equilibration
is with respect to the temperature.  A unified framework to study the isothermal class of ensembles
in introduced in this section. This section is divided into two subsections. In the first subsection
a unified formulation of the unnormalized $q$-expectation values or the second constraint formalism
is presented.   A similar unified method of the third constraint formalism is presented in the second
subsection.  Finally we demonstrate the interrelation between the two formalisms and derive a generalized
expression for the temperature transformation relation.
\subsection{Second constraint}
The expectation values of energy, a general thermodynamic observable and the heat function
based on the unnormalized $q$-expectation value (\ref{q_O_2}) is
\bea
U_{q}^{(2)} (\mathfrak{X}_{1},\mathfrak{X}_{2},\beta)
                          &=& \sum_{i,X_{\{\ell\}}}
                         \left({\mathfrak{p}}_{i:X_{\{\ell\}}}^{(2)}
                         ({\mathfrak{X}}_{1},
                         {\mathfrak{X}}_{2},\beta)\right)^{q}
                         \; \epsilon_{i},
\label{ie_2c_def}  \\
X_{\ell:q}^{(2)}(\mathfrak{X}_{1},\mathfrak{X}_{2},\beta)
                         &=& \sum_{i,X_{\{\ell\}}}
                         \left({\mathfrak{p}}_{i:X_{\{\ell\}}}^{(2)}
                         ({\mathfrak{X}}_{1},
                         {\mathfrak{X}}_{2},\beta)\right)^{q}
                         \; X_{\ell}.
\label{gq_2c_def} \\
\mathfrak{H}_{q}^{(2)} (\mathfrak{X}_{1},\mathfrak{X}_{2},\beta)
                          &=& \sum_{i,X_{\{\ell\}}}
                         \left({\mathfrak{p}}_{i:X_{\{\ell\}}}^{(2)}
                         ({\mathfrak{X}}_{1},
                         {\mathfrak{X}}_{2},\beta)\right)^{q}
                         \; \Big(\epsilon_{i} + \sum_{\{\ell\}}
                         x_{\ell} X_{\ell} \Big)
\label{hf_2c_def}
\eea
where ${\mathfrak{p}}_{i:X_{\{\ell\}}}^{(2)}$ is the probability of finding a particle in a
particular microstate and $X$ is an extensive thermodynamic quantity whose intensive counterpart is $x$.
The chemical and the mechanical equilibrium variables are denoted by
${\mathfrak{X}}_{1}$ and ${\mathfrak{X}}_{2}$. A brief introduction of the variable $\ell$
and the method to deduce the various ensembles corresponding to its different values is given in
Section \ref{Intro}.

\par

Optimization of the entropy (\ref{q_entr}) using the normalization
condition, the internal energy (\ref{ie_2c_def}) and the other
extensive thermodynamic variables (\ref{gq_2c_def}) leads to the
distribution given below
\beq
{\mathfrak{p}}_{i:X_{\{\ell\}}}^{(2)}({\mathfrak{X}}_{1},{\mathfrak{X}}_{2},\beta)
                          = \frac{1}{{\mathcal{Z}}_{q}^{(2)}
                            ({\mathfrak{X}}_{1},{\mathfrak{X}}_{2},\beta)}
                            \exp_{q} \bigg(-\beta
                            \Big(\epsilon_{i}+\sum_{\{\ell\}}x_{\ell}
                            X_{\ell}\Big)\bigg),
\label{pdf_2c}
\eeq
where ${\mathcal{Z}}_{q}^{(2)}$ the partition function in the second constraint is
\beq
{\mathcal{Z}}_{q}^{(2)}({\mathfrak{X}}_{1},{\mathfrak{X}}_{2},\beta) = \sum_{i,X_{\{\ell\}}}
                       \exp_{q} \bigg( -\beta
                       \Big(\epsilon_{i}+\sum_{\{\ell\}}x_{\ell}X_{\ell}\Big) \bigg).
\label{pf_ib_2c}
\eeq
As previously stated in section \ref{Intro}, we associate the temperature with the
Lagrange multiplier corresponding to the internal energy i.e., $\beta = 1/k T$.

\par

The second constraint heat function of a given ensemble
$({\mathfrak{X}}_{1},{\mathfrak{X}}_{2},\beta)$ can be computed from
the partition function using
\beq
{\mathfrak{H}}_{q}^{(2)}({\mathfrak{X}}_{1},{\mathfrak{X}}_{2},\beta)
                          = - \frac{\partial}{\partial \beta}
                             \ln_{q} {\mathcal{Z}}_{q}^{(2)}
                             ({\mathfrak{X}}_{1},{\mathfrak{X}}_{2},\beta).
\label{enth_def}
\eeq
In terms of the partition function the extensive variable in the
second constraint is
\beq
X_{\ell : q}^{(2)}({\mathfrak{X}}_{1},{\mathfrak{X}}_{2},\beta)
                            = - \frac{1}{\beta} \frac{\partial}
                              {\partial x_{\ell}}
                              \ln_{q}{\mathcal{Z}}_{q}^{(2)}
                              ({\mathfrak{X}}_{1},{\mathfrak{X}}_{2},\beta).
\label{av_2c_def}
\eeq
The knowledge of the heat function and the extensive thermodynamic
variables of a given ensemble enables us to find the internal energy
using Legendre transformation.

\par

The thermodynamic potential corresponding to an ensemble in the
second constraint formulation is the generalized free energy
\beq
{\mathcal{F}}_{q}^{(2)}
                      = - k T \ln_{q}{\mathcal{Z}}_{q}^{(2)}
                         ({\mathfrak{X}}_{1},{\mathfrak{X}}_{2},\beta),
\label{Gfe_2c}
\eeq
from which the entropy and the other extensive variables are computed as follows:
\beq
S_{q} = - \frac{\partial  {\mathcal{F}}_{q}^{(2)}}{\partial T}{\Bigg{|}}_{P},
\qquad
X_{\ell:q}^{(2)} = \frac{\partial  {\mathcal{F}}_{q}^{(2)}}{\partial x_{\ell}}
                         {\Bigg{|}}_{T}.
\label{av_2c_def1}
\eeq
The thermodynamic expression for the specific heats corresponding to the different conditions
is
\beq
C_{q}^{(2)}{\big{|}}_{V} = \frac{\partial  U_{q}^{(2)}}{\partial T}
                           {\Bigg{|}}_{V},
\qquad
C_{q}^{(2)}{\big{|}}_{x_{\{\ell\}}} = - T \, \frac{{\partial}^{2} {\mathcal{F}}_{q}^{(2)}}
                             {\partial T^{2}} {\Bigg{|}}_{x_{\{\ell\}}}
                         =   T \, \frac{\partial S_{q}}{\partial T}.
\label{psh_2c_def}
\eeq
\subsection{Third constraint formalism}
To overcome the failures of the second constraint we use the normalized
$q$-expectation values in the third constraint formalism. The definition
of the internal energy, an arbitrary extensive variable and the heat
function in this formalism are
\bea
U_{q}^{(3)}({\mathfrak{X}}_{1},{\mathfrak{X}}_{2},\beta) &=&
                         \frac{1}{\mathfrak{c}^{(3)}
                         ({\mathfrak{X}}_{1},{\mathfrak{X}}_{2},\beta)} \;
                         \sum_{i,X_{\{\ell\}}}
                         \left({\mathfrak{p}}_{i:X_{\{\ell\}}}^{(3)}
                         ({\mathfrak{X}}_{1},{\mathfrak{X}}_{2},\beta)\right)^{q} \;
                         \epsilon_{i},
\label{ge_ie_3c} \\
X_{\ell:q}^{(3)}({\mathfrak{X}}_{1},{\mathfrak{X}}_{2},\beta) &=&
                         \frac{1}{\mathfrak{c}^{(3)}
                         ({\mathfrak{X}}_{1},{\mathfrak{X}}_{2},\beta)} \;
                         \sum_{i,X_{\{\ell\}}}
                         \left({\mathfrak{p}}_{i:X_{\{\ell\}}}^{(3)}
                         ({\mathfrak{X}}_{1},{\mathfrak{X}}_{2},\beta)\right)^{q}
                         \; X_{\ell},
\label{ge_av_3c} \\
\mathfrak{H}_{q}^{(3)}({\mathfrak{X}}_{1},{\mathfrak{X}}_{2},\beta) &=&
                         \frac{1}{\mathfrak{c}^{(3)}
                         ({\mathfrak{X}}_{1},{\mathfrak{X}}_{2},\beta)} \;
                         \sum_{i,X_{\{\ell\}}}
                         \left({\mathfrak{p}}_{i:X_{\{\ell\}}}^{(3)}
                         ({\mathfrak{X}}_{1},{\mathfrak{X}}_{2},\beta)\right)^{q}
                          \Big(\epsilon_{i} + \sum_{\{\ell\}} x_{\ell} X_{\ell} \Big),
\label{hf_def_3c}
\eea
where $\mathfrak{p}_{i:X_{\{\ell\}}}^{(3)}$ is the probability of finding the particle in a particle
microstate. The factor ${\mathfrak{c}}^{(3)}$ is the sum of $q$-weights in the third constraint and
is defined as
\beq
\mathfrak{c}^{(3)}({\mathfrak{X}}_{1},{\mathfrak{X}}_{2},\beta,) =  \sum_{i,X_{\{\ell\}}} \left({\mathfrak{p}}_{i:X_{\{\ell\}}}^{(3)}
                                                                  ({\mathfrak{X}}_{1},{\mathfrak{X}}_{2},\beta)\right)^{q}.
\label{qw_esp_rel}
\eeq
Along with the normalization condition, the definitions corresponding
to the internal energy (\ref{ge_ie_3c}) and the arbitrary extensive
variable (\ref{ge_av_3c}) are used in the optimization of
the entropy (\ref{q_entr}). The probability distribution obtained through this
procedure is
\beq
{\mathfrak{p}}_{i:X_{\{\ell\}}}^{(3)}({\mathfrak{X}}_{1},{\mathfrak{X}}_{2},\beta) =
                                     \frac{1}{{\bar{{\mathcal{Z}}}_{q}}^{(3)}
                                     ({\mathfrak{X}}_{1},{\mathfrak{X}}_{2},\beta)} \;
                                     \exp_{q}\bigg(-\frac{\beta}
                                     {{\mathfrak{c}}^{(3)}
                                    ({\mathfrak{X}}_{1},{\mathfrak{X}}_{2},\beta)}
                                     \Big(\epsilon_{i} + \sum_{\{\ell\}}
                                     x_{\ell} X_{\ell} -
                                     {\mathfrak{H}}_{q}^{(3)}\Big)\bigg),
\label{pdf_3c}
\eeq
where ${\mathfrak{H}}_{q}^{(3)}$ is the heat function in the third constraint and $\beta = 1/k T$.
The generalized partition function ${\bar{\mathcal{Z}}}_{q}^{(3)}$ in
(\ref{pdf_3c}) is as defined below:
\beq
{{\bar{{\mathcal{Z}}}_{q}}^{(3)}(\beta,{\mathfrak{X}}_{1},{\mathfrak{X}}_{2}})
                        = \displaystyle \sum_{i,X_{\{\ell\}}}
                          \exp_{q} \bigg(-\frac{\beta}
                         {{\mathfrak{c}}^{(3)}(\beta,{\mathfrak{X}}_{1},
                          {\mathfrak{X}}_{2})}
                          \Big(\epsilon_{i} + \sum_{\{\ell\}}x_{\ell} X_{\ell}
                          - {\mathfrak{H}}_{q}^{(3)}\Big)\bigg).
\label{gpf_3c_def}
\eeq
The generalized partition function and the sum of the $q$-weights are related via the
expression
\beq
\left({{\bar{{\mathcal{Z}}}_{q}}^{(3)}
                    (\beta,{\mathfrak{X}}_{1},{\mathfrak{X}}_{2})}\right)^{1-q} =
                    {\mathfrak{c}}^{(3)}(\beta,{\mathfrak{X}}_{1},{\mathfrak{X}}_{2}),
\label{gpf_qw_rel}
\eeq
which holds for all the isothermal ensembles and plays a crucial role in
solving the implicit equations in the third constraint.

\par

The heat function in the third constraint satisfies the differential equation
\beq
\beta \, \frac{\partial}{\partial \beta}
{\mathfrak{H}}_{q}^{(3)}(\beta,{\mathfrak{X}}_{1},{\mathfrak{X}}_{2}) = \frac{\partial}
              {\partial \beta} \ln_{q} {\bar{\mathcal{Z}}_{q}}^{(3)}
              (\beta,{\mathfrak{X}}_{1},{\mathfrak{X}}_{2}).
\label{enth_def_3c}
\eeq
From the knowledge of the heat function ${\mathfrak{H}}_{q}^{(3)}$, and the
generalized partition function ${\bar{\mathcal{Z}}}_{q}^{(3)}$ the expectation
value of the extensive thermodynamic variable is found using
\beq
X_{\ell:q}^{(3)} (\beta,{\mathfrak{X}}_{1},{\mathfrak{X}}_{2})
                 =  \frac{\partial}{\partial x_{\ell}}
                    \left({\mathfrak{H}}_{q}^{(3)}
                    (\beta,{\mathfrak{X}}_{1},{\mathfrak{X}}_{2})
                    - \frac{1}{\beta}
                    \ln_{q} {\bar{\mathcal{Z}}_{q}}^{(3)}
                    (\beta,{\mathfrak{X}}_{1},{\mathfrak{X}}_{2})\right).
\label{avg_def_3c}
\eeq
The internal energy can be obtained from the heat function and the
extensive thermodynamic quantities using Legendre transformation.
Making use of the property (\ref{gpf_qw_rel}), the nonextensive entropy
(\ref{q_entr}) expressed in terms of the generalized partition
function (\ref{gpf_3c_def}) is
\beq
S_{q} = k \ln_{q}{\bar{\mathcal{Z}}}_{q}^{(3)}
        (\beta,{\mathfrak{X}}_{1},{\mathfrak{X}}_{2}).
\label{entropy_3c}
\eeq
Substituting the entropy (\ref{entropy_3c}) in (\ref{gfe_def}) we
arrive at the free energy in the third constraint as
\beq
{\mathcal{F}}_{q}^{(3)} = U_{q}^{(3)} - \frac{1}{\beta}
              \ln_{q} {\bar{\mathcal{Z}}}_{q}^{(3)}
              (\beta,{\mathfrak{X}}_{1},{\mathfrak{X}}_{2})
              + \sum_{\{\ell\}} x_{\ell} X_{\ell:q}^{(3)}.
\label{Gfe_3c_ls}
\eeq
In terms of the generalized free energy defined in
(\ref{Gfe_3c_ls}) the heat function and the $q$-expectation
value of the extensive thermodynamic variable reads:
\beq
{\mathfrak{H}}_{q}^{(3)} = \frac{\partial}{\partial \beta}
                           (\beta {\mathcal{F}}_{q}^{(3)}), \qquad
X_{\ell:q}^{(3)}  = \frac{\partial {\mathcal{F}}_{q}^{(3)}}
                    {\partial x_{\ell}}.
\label{Vol_hf_Gfe}
\eeq
The heat capacities at constant pressure and at constant volume
in the third constraint are
\bea
C_{q}^{(3)}{\Big{|}}_{V} &=& \frac{\partial U_{q}^{(3)}}{\partial T},
\label{vsh_3c_def}  \\
C_{q}^{(3)}{\Big{|}}_{x_{\{\ell\}}} &=& \frac{\partial {\mathfrak{H}}_{q}^{(3)}}{\partial T}
                          =   T \frac{\partial  S_{q}}{\partial T}
                          = - T \frac{{\partial}^{2}{\mathcal{F}}_{q}^{(3)}}
                             {\partial T^{2}}.
\label{psh_3c_def}
\eea

\par

The third constraint approach is the currently accepted formulation
of nonextensive statistical mechanics and problems which do not involve
simulations can be solved directly in this approach. In the case of
Montecarlo and moleculardynamics simulations direct use of the
third constraint is difficult.  To overcome this, the calculations
are initially performed in the second constraint and the results are
then transferred to the corresponding quantities in the third constraint.
The transformation relation for the canonical ensemble
was introduced in [\cite{TMP}]. Adopting a similar procedure for the
unified framework of isothermal ensembles, the probabilities in the
second and the third constraint formalisms (\ref{pdf_2c},\ref{pdf_3c})
are related as follows
\beq
{\mathfrak{p}}_{i:X_{\{\ell\}}}^{(3)}(\beta,{\mathfrak{X}}_{1},{\mathfrak{X}}_{2}) =
                      {\mathfrak{p}}_{i:X_{\{\ell\}}}^{(2)}
                      (\beta^{\prime},{\mathfrak{X}}_{1},{\mathfrak{X}}_{2})
                       = \exp_{q} \Big(-\beta^{\prime}
                         \Big(\epsilon_{i}+\displaystyle{\sum_{\{\ell\}}}
                         x_{\ell} X_{\ell}\Big)\Big)
                         \left({\mathcal{Z}}_{q}^{(2)}
                         \left(\beta^{\prime},{\mathfrak{X}}_{1},{\mathfrak{X}}_{2}\right)
                         \right)^{-1},
\label{pdf_rel}
\eeq
where, the pseudoinverse temperature $\beta^{\prime}$ is related to $\beta$ as
\beq
\beta^{\prime} = \frac{\beta} {{\mathfrak{c}}^{(3)}
                 (\beta,{\mathfrak{X}}_{1},{\mathfrak{X}}_{2})
                 + (1-q) \; \beta \; {\mathfrak{H}}_{q}^{(3)}
                (\beta,{\mathfrak{X}}_{1},{\mathfrak{X}}_{2})}.
\label{bpb_rel}
\eeq
The relation in (\ref{bpb_rel}) is inverted to obtain $\beta^{\prime}$ in
terms of $\beta$
\beq
\beta = {\beta}^{\prime}\frac{{\mathfrak{c}}^{(2)}
        (\beta^{\prime},{\mathfrak{X}}_{1},{\mathfrak{X}}_{2})}
        {1-(1-q)\, {\beta}^{\prime} \; \frac{{\mathfrak{H}}_{q}^{(2)}
        (\beta^{\prime},{\mathfrak{X}}_{1},{\mathfrak{X}}_{2})}
        {{\mathfrak{c}}^{(2)}(\beta^{\prime},{\mathfrak{X}}_{1},
        {\mathfrak{X}}_{2})}}.
\label{bbp_rel_1}
\eeq
After obtaining the transformation pertinent to a given system, there are two
independent methods through which we can get the other thermodynamic
variables.

\par

{\it(i)} Probability transformation technique:
In this method, introduced in Ref. [\cite{TMP}], the sum of $q$-weights
is transformed from the second constraint to the third constraint using the
temperature transformation relation (\ref{bbp_rel_1}).  Using
(\ref{gpf_qw_rel}) the generalized partition function is obtained from the
sum of $q$-weights from which we can calculate the specific heat directly.

\par

{\it(ii)} Observable transformation technique:
Any dynamical observable in the second constraint can be transformed to the
corresponding one in the third constraint [\cite{CCN2}] through the relation
given below
\beq
O_{q}^{(3)}(\beta,{\mathfrak{X}}_{1},{\mathfrak{X}}_{2})
         = \frac{O_{q}^{(2)} (\beta^{\prime},{\mathfrak{X}}_{1},
           {\mathfrak{X}}_{2})}
           {{\mathfrak{c}}^{(2)}(\beta^{\prime},{\mathfrak{X}}_{1},
           {\mathfrak{X}}_{2})}.
\label{Otrans}
\eeq
All the relevant thermodynamic quantities of a given ensemble in the third
constraint can be computed from the second constraint using the two methods
described above.

The thermodynamic relations for the canonical ensemble can be obtained
from the unified framework by observing that the internal energy and
the Helmholtz free energy are the heat function and the free energy
respectively. An elaborate exposition on the canonical ensemble may be
found in [\cite{CUR},\cite{TMP}] and since this ensemble has been
investigated in detail we do not intend to discuss it anymore.

%
%
%
%
\setcounter{equation}{0}
\section{Isothermal-Isobaric ensemble}
\label{isobaric}
The isothermal-isobaric ensemble is used to study systems which attain
equilibrium by exchanging both internal energy and volume with the heat bath.
The enthalpy and the Gibbs free energy plays the role of heat function and free energy
respectively. Using the unified framework described in Section \ref{Isothermal} the
nonrelativisitic and the extreme relativistic ideal gases are investigated.
\subsection{Nonrelativistic Ideal gas}
\label{CIG}
The Hamiltonian of a classical ideal gas in the non-relativistic
regime with particles possessing momenta ${\bf{p}}_{i}$ $(i=1,...,N)$
in $D$ dimensions is
\beq
{\mathcal{H}}_{nr}(p) = \sum_{i} \frac{p_{i}^{2}}{2m},
\qquad \qquad
p_{i}=|{\bf{p}}_{i}|.
\label{Ham_nr}
\eeq
The generalized partition function corresponding to the third constraint
in the $(N,P,T)$ ensemble is
\bea
{\bar{Y}}_{q}^{(3)}(\beta,P) &=&   \frac{1}{N! h^{DN}}
                                   \displaystyle \int_{V} {\rm d}V
                                   \int_{r_{i}}\int_{p_{i}}
                                   \prod_{i=1}^{N} {\rm d}^{D}r_{i}
                                   {\rm d}^{D}p_{i} \; \times \nn \\
                             & &   \times \; \exp_{q} \left(- \frac{\beta}
                                   {\mathfrak{c}^{(3)}}
                                   \left(\displaystyle \sum_{i}
                                   \frac{p_{i}^{2}}{2m}-U_{q}^{(3)}
                                   +PV-P {V}_{q}^{(3)} \right)\right).
\label{gpf_2c_int}
\eea
Carrying out the phase space integration in (\ref{gpf_2c_int}), we arrive
at
\beq
{\bar{Y}}_{q}^{(3)}(\beta,P) = \frac{\mathcal{M}^{N}}{N!} \,
                               \frac{\Gamma \left(\frac{2-q}{1-q}\right)}
                               {\Gamma \left(\frac{2-q}{1-q}+\frac{DN}{2}
                               \right)}\int_{V} V^{N} \left(
                               \exp_{q}\left(\frac{\beta}{c^{(3)}}
                               \left(U_{q}^{(3)}-PV+PV_{q}^{(3)}\right)
                               \right) \right)^{\Lambda} {\rm d} V,
\label{gpf_mi_3c}
\eeq
where ${\mathcal{M}} = \left(\frac{2 \pi m}{h^{2}}\right)^{D/2}$
and $\Lambda = 1+(1-q)DN/2$. In order to express the
thermodynamic variables in a simplified manner, we define the following
quantities
\beq
\Lambda_{\kappa} = 1+ (1-q) {{\mathfrak{D}}_{\kappa}}, \quad
{\mathfrak{L}}_{\kappa} =  (1-(1-q){{\mathfrak{D}}_{\kappa}})^{-1}, \quad
\mu_{\kappa} = \Lambda_{\kappa} {\mathfrak{L}}_{\kappa},
\label{Lam_mfl_def}
\eeq
where $\kappa = 1 \, (2)$ for a
non-relativistic (extreme relativistic) case and
${\mathfrak{D}}_{1}=\frac{DN}{2}+N$ $({\mathfrak{D}}_{2}=DN+N)$.
Integrating over volume in (\ref{gpf_mi_3c}) we arrive at the
generalized partition function
\bea
{\bar{Y}}_{q}^{(3)}(\beta,P) &=& \frac{\mathcal{M}^{N} \;\Gamma\left(\frac{2-q}
                                   {1-q}\right)}{\Gamma \left(\frac{2-q}{1-q}
                                  +{\mathfrak{D}}_{1}+1 \right)} \left(\frac{1}
                                  {1-q}\right)^{{\mathfrak{D}}_{1}+1}
                                   \left( \exp_{q} \left(\frac{\beta}
                                   {{\mathfrak{c}}^{(3)}}
                                   (U_{q}^{(3)} + P V_{q}^{(3)})\right)\right)
                                   ^{{\Lambda_{1}}-(1-q)}
                                   \nn \\
                              & & \left(\frac{{\mathfrak{c}}^{(3)}}
                                   {\beta}\right)^{{\mathfrak{D}}_{1}+1}
                                   \left(\frac{1}{P}\right)^{N+1}.
\label{gpf_vdim}
\eea
It is interesting to note that (\ref{gpf_vdim}) has the dimensions of volume.
This is due to the overcounting of volume eigenstates in systems in which the
volume is a continuous variable. To overcome this we use the shell particle
method of counting of volume states [\cite{DC1},\cite{DC2}]. In the shell
particle method of counting we take into account only the minimum volume needed to
confine a particular configuration.  When there are many equivalent ways
of choosing a minimum volume we treat all of them as the same volume
eigenstate. The minimum volume is fixed by enforcing a condition where
we require that at least one particle lies on the boundary of the system.
Carrying out the volume integral of the generalized partition function using the
shell particle counting technique to reject the redundant volume states we arrive at
\beq
{\bar{Y}}_{q}^{(3)}(\beta,P) =  \mathcal{M}^{N} \; {\mathcal{G}}_{1}
                                   \left( \exp_{q} \left(\frac{\beta}
                                   {{\mathfrak{c}}^{(3)}}
                                   (U_{q}^{(3)} + P V_{q}^{(3)}) \right)\right)
                                   ^{\Lambda_{1}}
                                   \left(\frac{{\mathfrak{c}}^{(3)}}
                                   {\beta}\right)^{{\mathfrak{D}}_{1}}
                                   \left(\frac{1}{P}\right)^{N}.
\label{gpf_2c_fs}
\eeq
The above expression of the generalized partition function (\ref{gpf_2c_fs}) in contrast
to (\ref{gpf_vdim}), is dimensionless making it consistent with the basic
principles of statistical mechanics. The ratio of gamma functions
${\mathcal{G}}_{1}$ made use of in (\ref{gpf_2c_fs}) is
\beq
{\mathcal{G}}_{1} = \frac{\Gamma \big( \frac{2-q}{1-q} \big)}
                {(1-q)^{{\mathfrak{D}}_{1}} \;
                \Gamma \big( \frac{2-q}{1-q}+{\mathfrak{D}}_{1} \big)}.
\label{cM_cG_def}
\eeq

\par

The heat function of the isothermal-isobaric ensemble given by the basic
relation (\ref{hf_def_3c}) is computed and the result is
\beq
H_{q}^{(3)} = {\mathfrak{D}}_{1} \; \frac{\mathcal{M}^{N}}{{\bar{Y}}_{q}^{(3)}} \;\;
              {\mathcal{G}}_{1} \;\;
              \left( \exp_{q} \left(\frac{\beta}
              {{\mathfrak{c}}^{(3)}} H_{q}^{(3)} \right)\right)^{\Lambda_{1}} \;
              \left(\frac{1}{P}\right)^{N} \;
              \left(\frac{{\mathfrak{c}}^{(3)}}{\beta}\right)^{\mathfrak{D}_{1}+1}.
\label{en_imp_3c}
\eeq
The ratio between the (\ref{gpf_2c_fs}) and (\ref{en_imp_3c}) is computed
\beq
H_{q}^{(3)} = {\mathfrak{D}}_{1}  \;  \frac{{\mathfrak{c}}^{(3)}}{\beta}.
\label{en_qw_rel}
\eeq
Using (\ref{en_qw_rel}) in conjunction with (\ref{gpf_qw_rel}), in (\ref{gpf_2c_fs}) the explicit
form of the generalized partition function is
\beq
{\bar{Y}}_{q}^{(3)} = \left(\mathcal{M}^{N} \; {\mathcal{G}}_{1}\right)
                      ^{{\mathfrak{L}}_{1}}
                      \left(\exp_{q} ({\mathfrak{D}}_{1})\right)
                      ^{\mu_{1}} \;
                      \left(\frac{1}{\beta}\right)
                      ^{{\mathfrak{D}}_{1} {\mathfrak{L}}_{1}}
                      \left(\frac{1} {P}\right)
                      ^{N {\mathfrak{L}}_{1}}.
\label{gpf_3c_fe}
\eeq
Similarly the explicit expression of the enthalpy computed through the use of
(\ref{en_imp_3c}), (\ref{gpf_3c_fe}) and (\ref{gpf_qw_rel}) is
\beq
H_{q}^{(3)} = \frac{{\mathfrak{D}}_{1}}{\beta} \;\; \Xi_{nr} \;
              \bigg( \frac{1}{\beta} \bigg)
              ^{(1-q){\mathfrak{D}}_{1}{\mathfrak{L}}_{1}}
              \bigg( \frac{1}{P} \bigg)
              ^{(1-q) N {\mathfrak{L}}_{1}},
\label{en_3c_fe}
\eeq
where the constant factor $\Xi_{nr}$ used in (\ref{en_3c_fe}) is
\beq
\Xi_{nr} = \left(\mathcal{M}^{N} \; {\mathcal{G}}_{1}\right)
           ^{(1-q){\mathfrak{L}}_{1}} \; \Lambda_{1}^{\mu_{1}}.
\label{xi_def}
\eeq
The entropy of the system in the third constraint written down from the
generalized partition function using the relation (\ref{entropy_3c}) is
\beq
S_{q} = \frac{k}{1-q} \s{ \Xi_{nr} \; \left( \frac{1}{\beta} \right)
        ^{(1-q) {{\mathfrak{D}}_{1}} {\mathfrak{L}}_{1}}
        \left(\frac{1}{P}\right)
        ^{(1-q) N {\mathfrak{L}}_{1}} -1}.
\label{ent_3c_fin}
\eeq
Making use of the thermodynamic Legendre structure (\ref{Gfe_3c_ls})
the Gibbs free energy of the classical ideal gas is obtained:
\beq
G_{q}^{(3)} =  \frac{1}{(1-q) \beta} \s{ ((1-q){\mathfrak{D}}_{1}-1) \;
                \Xi_{nr} \; \left( \frac{1}{\beta} \right)
                ^{(1-q){{\mathfrak{D}}_{1}}
                {\mathfrak{L}}_{1}}
                \left( \frac{1}{P} \right)
                ^{(1-q) N {\mathfrak{L}}_{1}} -1}.
\label{gfe_3c_fin}
\eeq

\par

The average volume of the classical nonrelativistic ideal gas in the third constraint
can be found via the expression (\ref{avg_def_3c}) using the explicit expression of the
generalized partition function (\ref{gpf_3c_fe}) and the enthalpy (\ref{en_3c_fe}).
The calculated expression of the average volume reads:
\beq
V_{q}^{(3)} = \frac{N}{\beta \, P} \; \Xi_{nr} \;
                \bigg( \frac{1}{\beta} \bigg)
                ^{(1-q){\mathfrak{D}}_{1}{\mathfrak{L}}_{1}}
                \bigg( \frac{1}{P} \bigg)
                ^{(1-q) N {\mathfrak{L}}_{1}}.
\label{av_3c_fe}
\eeq
The average volume which was computed using enthalpy can also be calculated through two other
different methods. The first method is based on a specific form of (\ref{Vol_hf_Gfe}) applicable
to the isothermal-isobaric ensemble in which the average volume is related to the Gibbs free
energy. The second method makes use of the defining relation of the average volume (\ref{ge_av_3c}) to
obtain an implicit expression. Using the implicit expression in conjunction with (\ref{en_3c_fe}),
(\ref{gpf_3c_fe}) and (\ref{gpf_qw_rel}) we can obtain the final explicit form. The results obtained through
both these methods are in agreement with that obtained in (\ref{av_3c_fe}). Rewriting the expression for the
average volume (\ref{av_3c_fe}) we arrive at the nonextensive form of the equation of state corresponding
to the isothermal-isobaric ensemble
\beq
P V_{q}^{(3)} = \frac{N}{\beta} \;\;
                        \Xi_{nr} \; \left(\frac{1}{\beta}\right)
                       ^{(1-q){\mathfrak{D}}_{1} {\mathfrak{L}}_{1}}
                       \left(\frac{1}{P}\right)
                       ^{(1-q) N {\mathfrak{L}}_{1}}.
\label{eq_st_3c}
\eeq
From the knowledge of the enthalpy and the average volume, the internal
energy evaluated using the Legendre structure is
\beq
U_{q}^{(3)} = \frac{DN}{2 \beta} \; \Xi_{nr} \;
                \bigg( \frac{1}{\beta} \bigg)
                ^{(1-q){\mathfrak{D}}_{1}{\mathfrak{L}}_{1}}
                \bigg( \frac{1}{P} \bigg)
                ^{(1-q) N {\mathfrak{L}}_{1}}.
\label{ie_3c_fe}
\eeq
The defining relation of the internal energy (\ref{ge_ie_3c}) has been used to verify (\ref{ie_3c_fe})
in a manner analogous to the calculation of the average volume.

\par

The specific heat at constant pressure can be found from either the
enthalpy (\ref{en_3c_fe}) or entropy (\ref{ent_3c_fin}) or the
Gibbs free energy (\ref{gfe_3c_fin}) using (\ref{psh_3c_def})
\beq
C_{q}^{(3)} \big|_{P} = \left(\frac{DN}{2} + N \right) k \;\;
                        {\mathfrak{L}}_{1} \; \Xi_{nr} \;
                        \left(\frac{1}{\beta}\right)
                        ^{(1-q){{\mathfrak{D}}_{1}} {\mathfrak{L}}_{1}}
                        \left(\frac{1} {P} \right)
                        ^{(1-q) N {\mathfrak{L}}_{1}}.
\label{psh_3c_fin}
\eeq
From the internal energy (\ref{ie_3c_fe}) the specific heat at constant
volume is evaluated through the use of (\ref{vsh_3c_def})
\beq
C_{q}^{(3)} \big|_{V} = \frac{DN}{2} \ k \;\; \Xi_{nr} \;\;{\mathfrak{L}}_{1} \;
                        \left(\frac{1}{\beta}\right)
                        ^{(1-q){{\mathfrak{D}}_{1}} {\mathfrak{L}}_{1}}
                        \left(\frac{1} {P} \right)
                        ^{(1-q) N {\mathfrak{L}}_{1}}.
\label{vsh_3c_fin}
\eeq
The difference between the specific heat at constant pressure (\ref{psh_3c_fin}) and the
specific heat at constant volume (\ref{vsh_3c_fin}) is
\beq
C_{q}^{(3)} \big|_{P} - C_{q}^{(3)} \big|_{V} = N k \;\;
                                                {\mathfrak{L}}_{1} \;
                                                \Xi_{nr} \;
                                                \left(\frac{1}{\beta}\right)
                                                ^{(1-q){{\mathfrak{D}}_{1}}
                                                {\mathfrak{L}}_{1}}
                                                \left(\frac{1} {P}\right)
                                                ^{(1-q) N
                                                {\mathfrak{L}}_{1}},
\label{May_rel_3c}
\eeq
which is the generalized form of Meyer's relation and in the extensive limit
the lhs goes as $Nk$.
It is interesting to note that the factor $\gamma$ expressing the ratio
between the two specific heats
\beq
\gamma = \frac{C_{q}^{(3)} \big|_{P}} {C_{q}^{(3)} \big|_{V}}
       = 1 + \frac{2}{D},
\label{poly_ind}
\eeq
is independent of the nonextensive parameter $q$ and the number of particles $N$, though the
specific heats (\ref{psh_3c_fin}) and (\ref{vsh_3c_fin}) depend on these parameters explicitly.

\par

The thermodynamic quantities calculated in this section can
also be evaluated in a different method where we compute the thermodynamic
quantities in the second constraint framework. The ensemble probabilities
in the second and the third constraints are related through a fictitious
temperature $\beta^{\prime}$ as given in equation (\ref{pdf_rel}). This
fictitious temperature $\beta^{\prime}$ in the second constraint is related to
the original temperature $\beta$ in the third constraint through a
transformation. The temperature transformation relation corresponding to the
classical ideal gas in the isothermal-isobaric ensemble is
\beq
\frac{1}{\beta^{\prime}} = (\mathcal{M}^{N} \; {\mathcal{G}}_{1})
                           ^{(1-q) {\mathfrak{L}}_{1}} \;
                           \Lambda_{1}^{2 {\mathfrak{L}}_{1}} \;
                           \left(\frac{1}{P}\right)
                           ^{(1-q) N {\mathfrak{L}}_{1}}
                           \left(\frac{1}{\beta}\right)
                           ^{{\mathfrak{L}}_{1}}.
\label{bbp_nrg}
\eeq
Using the above transformation the third constraint thermodynamic quantities
can be computed using the probability transformation technique
or the observable transformation technique. The thermodynamic quantities
and the equation of state obtained through this transformation method are
identical to the corresponding quantities computed directly in the third
constraint.
\subsection{Extreme relativistic ideal gas}
The Hamiltonian of the extreme relativistic classical ideal gas in
$D$ dimensions is
\beq
{\mathcal{H}}_{er}(p) = c \; \sum_{i} p_{i},
\qquad \qquad
p_{i}=|{\bf{p}}_{i}|,
\label{Ham_er}
\eeq
where the factor $c$ is the velocity of light. The nonextensive study of the Hamiltonian
(\ref{Ham_er}) is carried out in the isothermal-isobaric ensemble.  Since the calculational
procedure is identical to the nonrelativistic case we present only the core steps leading to
the calculation of the specific heats. The generalized partition function calculated using
the basic definition is
\beq
\bar{Y}_{q}^{(3)} = \Delta^{N} \; \mathcal{G}_{2} \; \left(\frac{1}{P}\right)^{N}
                    \left(\frac{\mathfrak{c}^{(3)}}{\beta}\right)^{DN+N}.
\label{genpart_exrgas}
\eeq
The notations made use of in (\ref{genpart_exrgas})
are
\beq
\Delta = \frac{2 \pi^{D/2} \Gamma(D)}{(c h)^D \Gamma(D/2)},
\qquad \qquad
{\mathcal{G}}_{2} = \frac{\Gamma \left( \frac{2-q}{1-q} \right) }
                    {(1-q)^{{\mathfrak{D}}_{2}} \;
                    \Gamma \left( \frac{2-q}{1-q} + {\mathfrak{D}}_{2}
                    \right)}
\label{del_g2_def}
\eeq
Equation (\ref{genpart_exrgas}) is an implicit equation since it is a function of
the sum of $q$-weights which is related to the generalized partition function through
the relation (\ref{gpf_qw_rel}).  The nonextensive enthalpy of the extreme relativistic ideal
gas is
\beq
H_{q}^{(3)} = \mathfrak{D}_{2} \; \frac{\Delta^{N}}{\bar{Y}_{q}^{(3)}} \; \mathcal{G}_{2} \; \left(\exp_{q}\left(\frac{\beta}{\mathfrak{c}^{(3)}} H_{q}^{(3)} \right)\right)^{\Lambda_{2}} \left(\frac{1}{P}\right)^{N} \left(\frac{\mathfrak{c^{(3)}}}{\beta}\right)^{\mathfrak{D}_{2}+1}
\label{enth_exg_imp}
\eeq
where the factors $\Lambda_{2}$, ${\mathfrak{L}}_{2}$ and $\mu_{2}$ used in this section are described
in (\ref{Lam_mfl_def}) and the discussions following it. The ratio between (\ref{genpart_exrgas})
and (\ref{enth_exg_imp}) is
\beq
H_{q}^{(3)} = \mathfrak{D}_{2} \; \frac{\mathfrak{c}^{(3)}}{\beta}.
\label{enth_exg_exp}
\eeq
Using (\ref{enth_exg_exp}) in (\ref{genpart_exrgas}) we obtain the explicit expression corresponding to
the generalized partition function
\beq
\bar{Y}_{q}^{(3)} = (\Delta^{N}\mathcal{G}_{2})^{\mathfrak{L}_{2}} \;
                    \left(\exp_{q}(\mathfrak{D_{2}})\right)^{\mu_{2}}
                    \left(\frac{1}{P}\right)^{N \mathfrak{L}_{2}}
                    \left(\frac{1}{\beta}\right)^{\mathfrak{D}_{2}\mathfrak{L}_{2}}.
\label{genp_exp}
\eeq
Similarly the calculated value of the explicit form of the enthalpy is
\beq
H_{q}^{(3)}=\frac{\mathfrak{D}_{2}}{\beta} \; \Xi_{er}\; \left(\frac{1}{P}\right)^{(1-q)N \mathfrak{L}_{2}} \left(\frac{1}{\beta}\right)^{(1-q)\mathfrak{D}_{2}\mathfrak{L}_{2}},
\label{enth_exp}
\eeq
where the factor $\Xi_{er}$ is defined as
\beq
\Xi_{er} = \left(\Delta^{N}  \;
            {\mathcal{G}}_{2}\right)^{(1-q) {\mathfrak{L}}_{2}} \;
            \Lambda_{2}^{\mu_{2}}.
\eeq
The average volume of the extreme relativistic ideal gas is evaluated through the
use of (\ref{vol_N_def}). The equation of state obtained by rewriting the
expression for the average volume is
\beq
P V_{q}^{(3)} = \frac{N}{\beta} \;\;
                        \Xi_{er} \; \left(\frac{1}{\beta}\right)
                       ^{(1-q){\mathfrak{D}}_{2} {\mathfrak{L}}_{2}}
                       \left(\frac{1}{P}\right)
                       ^{(1-q) N {\mathfrak{L}}_{2}}.
\label{eq_st_rel_1}
\eeq
To find the internal energy we use the Legendre transform (\ref{hf_def}) and the
expressions for the enthalpy and the average volume.  The computed expression of
the internal energy reads:
\beq
U_{q}^{(3)} = \frac{DN}{\beta} \; \Xi_{er} \;
                \bigg( \frac{1}{\beta} \bigg)
                ^{(1-q){\mathfrak{D}}_{2}{\mathfrak{L}}_{2}}
                \bigg( \frac{1}{P} \bigg)
                ^{(1-q) N {\mathfrak{L}}_{2}}.
\label{ie_erg_fe}
\eeq
The specific heat capacity of the extreme relativistic ideal gas at constant volume
and at constant pressure can be established from the final explicit expressions of
the internal energy and the enthalpy respectively.  We exhibit the results below
\bea
C_{q}^{(3)} \big|_{V} &=& D N \;k \;\; \Xi_{er} \;\;{\mathfrak{L}}_{2} \;
                          \bigg(\frac{1}{\beta}\bigg)
                          ^{(1-q){{\mathfrak{D}}_{2}} {\mathfrak{L}}_{2}}
                          \bigg( \frac{1} {P} \bigg)
                          ^{(1-q) N {\mathfrak{L}}_{2}},
\label{vsh_rel_fin_1} \\
C_{q}^{(3)} \big|_{P} &=& (D N +N) \; k \;\; \Xi_{er} \;
                          {\mathfrak{L}}_{2} \;
                          \left(\frac{1}{\beta}\right)
                          ^{(1-q){{\mathfrak{D}}_{2}} {\mathfrak{L}}_{2}}
                          \left(\frac{1} {P}\right)
                          ^{(1-q) N {\mathfrak{L}}_{2}}.
\label{psh_rel_fin_1}
\eea
The ratio between the heat capacity at constant pressure and at constant volume
\beq
\frac{C_{q}^{(3)} \big|_{P}}{C_{q}^{(3)} \big|_{V}} = 1 + \frac{1}{D},
\label{sh_rat_erg}
\eeq
is found to be independent of both the nonextensive parameter $q$ and the number of particles $N$.
All the results in this section have been calculated through two methods. In the first method,
the third constraint generalized partition function, sum of $q$-weights and the enthalpy
are computed.  Since these quantities are implicit in nature, they are systematically solved
to obtain the explicit forms.  The second method is based on the interrelation between
the second and the third constraint formalisms.  The thermodynamic quantities are first
obtained in the second constraint formalism and they are later transformed to the corresponding
quantities in the third constraint through the use of the temperature transformation relation.
Both the probability transformation and the observable transformation techniques are used in
this objective. The results obtained through these methods are in precise agreement with each other.
The proper counting of volume states advocated in references [\cite{DC1},\cite{DC2}] has been used in both the methods. In the extensive $q \rightarrow 1$ limit, the standard physical quantities corresponding to the classical Boltzmann-Gibbs statistics are recovered.

\par

Another interesting model is the Tonks gas [\cite{LT1}] which is a one dimensional model of hard rod gas.
The Hamiltonian describing the dynamics of a Tonks gas is
\beq
{\mathcal{H}}_{tg}(p)  = \sum_{i} \frac{p_{i}^{2}}{2m} + V(r),
\qquad \qquad
p_{i}=|{\bf{p}}_{i}|,
\label{ham_tg}
\eeq
where the potential $V(r)$ is
\bea
V(r) &=& 0     \;  \qquad   x \geq \sigma
\label{tg_con}  \\
     &=& \infty  \qquad   x < \sigma.   \nn
\eea
An exact solution of the Tonks gas in the free length ensemble can be read from the
results of the nonrelativistic ideal gas.  In this discussion the term free length
refers to the actual volume available for a particle $L_{f:q} = L - (N-1)\sigma$.
The equation of state of the Tonks gas in the free length ensemble can
be read from the expression corresponding to the classical nonrelativistic ideal gas by
observing that the number of dimensions is $1$ and replacing the volume by the free length.
The equation of state of the Tonks gas in the free length ensemble is
\beq
P L_{f:q}^{(3)} = \frac{N}{\beta} \; \left(\mathcal{M}^{N}_{1}
                  \;{\mathcal{G}}_{1:1}\right)^{(1-q){\mathfrak{L}}_{1:1}} \;\;
                  (\Lambda_{1:1})^{\mu_{1:1}} \;  \left(\frac{1}{\beta}\right)
                  ^{\frac{3N}{2}{\mathfrak{L}}_{1:1}} \;\;
                  \left(\frac{1}{P}\right)^{N {\mathfrak{L}}_{1:1}},
\label{TG_Eos}
\eeq
where the functions ${\mathcal{G}}_{1:1}$  and $\Lambda_{1:1}$ are special cases
of (\ref{cM_cG_def}), obtained by substituting $D = 1$ and ${\mathcal{M}}_{1}$ is the
$1$ dimensional analog of $\mathcal{M}$.  A direct computation of the Tonks gas is used
to verify the expression for equation of state. The correct counting of volume states has
been applied and the Boltzmann-Gibbs results are recovered in the $q \rightarrow 1$ limit.
%
%
%
%
%
\setcounter{equation}{0}
\section{Grand Canonical ensemble}
\label{GCE}
A system which exchanges both the internal energy and the number of particles with its
surroundings is described using grand canonical ensemble. The Hill energy
$\mathsf{L} = U - \mu N$ is the heat function and the free energy is
$\Phi = U - T S - \mu N$. In this section we extend the perturbative formulation
developed in [\cite{CCN1}] and [\cite{CCN2}] to study the nonrelativistic and
extreme relativistic ideal gas in grandcanonical ensemble.
\subsection{Nonrelativistic classical ideal gas}
A perturbative study of the nonrelativistic classical ideal gas described by the
Hamiltoninan (\ref{Ham_nr}) is carried out in the grand canonical ensemble.  The
generalized partition function in the third constraint is
\beq
{\bar{Z}}_{q}^{(3)}(\mu,V,\beta) = \sum_{N=0}^{\infty} \frac{1}{N!\, h^{DN}}\int
                                   {\rm d}^{DN} r  \; \;  {\rm d}^{DN} p \;\;
                                   \exp_{q} \left(-{\tilde{\beta}} \left(\sum_{i}
                                   \frac{p_{i}^{2}}{2m} - \mu N - {\mathsf{L}}_{q}^{(3)}\right) \right),
\label{gpf_gc_def}
\eeq
where ${\tilde{\beta}} = \frac{\beta}{{\mathfrak{c}}^{(3)}}$ and ${\mathfrak{c}}^{(3)}$ is the
sum of $q$ weights corresponding to the grand canonical ensemble in the third constraint.
The Hill function ${\mathsf{L}}_{q}^{(3)} = U_{q}^{(3)} - \mu \, N_{q}^{(3)}$ is the heat function
corresponding to the grand canonical ensemble where, $U_{q}^{(3)}$ and $N_{q}^{(3)}$ are the
internal energy and the average number of particles respectively in the  third constraint and $\mu$ is the
chemical potential of the system.

\par

The observation in [\cite{CQ}], that the $q$ exponential can be written as an infinite multiplicative
series of ordinary exponential
\beq
\exp_{q}\left(x\right) = \prod_{k=1}^{\infty} \exp\left((-1)^{k-1} \frac{(1-q)^{k-1} \; x^{k}}{k}\right),
\label{qexp_clexp_rel}
\eeq
which in turn can be expanded up to any order in $(1-q)$, is used to construct a perturbative
procedure for the generalized partition function. Assuming the generalized partition function (\ref{gpf_gc_def})
to be well defined in the region $q=1$, we replace the $q$-exponential by the ordinary exponential.  The resultant expression for the generalized partition function is:
\beq
{\bar{Z}}_{q}^{(3)}(\mu,V,\beta) =   \widehat{{\mathcal{D}}}_{1}(d_{\beta}) \;
                                     \exp(\u)
                                     \sum_{N=0}^{\infty} \frac{1}{N!\, h^{DN}} \int
                                       {\rm d}^{DN} r  \;  {\rm d}^{DN} p \;
                                       \exp\bigg(-{\tilde{\beta}}\Big(\sum_{i}
                                       \frac{p_{i}^{2}}{2m} - \mu N\Big)\bigg).
\label{gpf_ps_gc}
\eeq
where $\u = {\tilde{\beta}} \, {\mathsf{L}}_{q}^{(3)}$ and the operator valued series
$\widehat{{\mathcal{D}}}_{1}(d_{\beta})$  in (\ref{gpf_ps_gc}) is
\beq
\widehat{{\mathcal{D}}}_{1}(d_{\beta}) = 1 - \frac{(1-q)}{2} \, d_{\beta}^{(2)} + \frac{(1-q)^{2}}{3} \;
                                         \left(d_{\beta}^{(3)} + \frac{3}{8} \; d_{\beta}^{(4)}\right)
                                         + \cdots ,
                                         \qquad
d_{\beta}^{(n)} = \beta^{n} \frac{\partial^{n}}{\partial \beta^{n}}.
\label{gpf_deriv_ser}
\eeq
In the expression (\ref{gpf_ps_gc}), we assume that the summation over the number of particles is a convergent,
atleast in the region where the nonextensive parameter is close to $1$.
Carrying out the phase space integration and the summation over the number of particles in (\ref{gpf_ps_gc}),
the generalized partition function in the third constraint can be expressed as
\beq
{\bar{Z}}_{q}^{(3)}(\mu,V,\beta) =  \widehat{{\mathcal{D}}}_{1}(d_{\beta}) \; \exp(\u) \;
                                    Z_{BG}({\tilde{\beta}}),
\label{gpf_BGpf_rel}
\eeq
where $Z_{BG}(\beta) = \exp\left(\exp(\beta \mu) \, V  \,{\mathcal{M}} \, \beta^{-D/2}\right)$
is the Boltzmann-Gibbs partition function.
The perturbative series of the generalized partition function (\ref{gpf_BGpf_rel}) calculated upto
second order in the expansion parameter $(1-q)$ is
\bea
{{\bar{{\mathcal{Z}}}_{q}}^{(3)}} = \exp(\u) \;  Z_{BG}(\tilde{\beta}) \;
                                    (1 + (1-q) \; Z_{1} + (1-q)^{2} \; Z_{2} +\cdots),
\label{gen_par_pert}
\eea
where the coeffecients of the perturbative series (\ref{gen_par_pert}) are listed below
\bea
Z_{1} &=& \frac{1}{4}
          \Big(2 \, \u^{2} + 2 N_{BG}({\tilde{\beta}}) \; (\a)^{2} + (N_{BG}({\tilde{\beta}}))^{2}\;
          \big(2 \, (\a)^{2} + 4 \,\a \; \u + D\big)\Big),\nn \\
Z_{2} &=&   \frac{1}{8}\,\u^{4} +\frac{1}{3} \,
            \u^{3} + \frac{1}{8} \, (N_{BG}({\tilde{\beta}}))^{4} \, (\a)^{4}
            + \frac{1}{96} \, (N_{BG}({\tilde{\beta}}))^{3} (72  \,(\a)^{4} + 12 \, \u \,(\a)^{3} \nn \\
      & &   + 4(8 \, \ab^{3}-3D^{2}) \, \a + D \,(4  \,\ab^{3} - D^{2}))
            + \frac{1}{32} \, (N_{BG}({\tilde{\beta}}))^{2} \,(48 \, \u \, (\a)^{2}\nn \\
      & &   + 8 \, \u \,(3 \,\u + 2) (\a)^{2} - 12 D \, {\tilde{\beta}} \mu \, \a + 8 \,(4-7D)\ab^{3} + 42 \, D^{2} \,\ab^{2}\nn\\
      & &   - 2D^{2} \,(9+7D) \, {\tilde{\beta}} + D^{2}(3+5D+7D^{2}))
            + \frac{1}{96} \, N_{BG}({\tilde{\beta}}) \,\Big(12 \,(\a)^{4}\nn \\
      & &   + 48 \, \u \, (\a)^{3} + (72 \, \u - 48) \, \u \, (\a)^{2} + (48 \, \u^{3}+72 \, \u^{2} - 12 \,D^{2})\a\nn \\
      & &   +64 \, \ab^{3} - 12D \,\ab^{2} + 24D \, {\tilde{\beta}}\mu  \, \u - D(4+9D-D^{2})\Big),
\label{gen_impl}
\eea
The factor $\alpha(\tilde{\beta}) = \tilde{\beta} - \frac{D}{2}$ and the average number in the standard Boltzmann-Gibbs
statistics is $N_{BG}(\beta) = \exp(\beta \mu) \, V  \,{\mathcal{M}} \, \beta^{-D/2}$ are used in (\ref{gen_impl}).
We notice that the generalized partition function described above is an implicit quantity and depends on the
heat function ${\mathsf{L}}_{q}^{(3)}$ and the sum of $q$-weights ${\mathfrak{c}}^{(3)}$ which is  in turn
related to the generalized partition function via the relation (\ref{gpf_qw_rel}).

\par

The integral form of the sum of $q$-weights written down from the defining relation (\ref{qw_esp_rel}) is
\bea
{\mathfrak{c}}^{(3)}(\mu,V,\beta) &=& \frac{1} {\left({\bar{Z}}_{q}^{(3)}(\mu,V,\beta)\right)^{q}} \; \times \\
                                  & & \times \, \sum_{N=0}^{\infty} \frac{1}{N!\, h^{DN}}
                                      \int {\rm d}^{DN} r  \; {\rm d}^{DN} p \;
                                      \left(\exp_{q}\bigg(-{\tilde{\beta}} \Big(\sum_{i}
                                      \frac{p_{i}^{2}}{2m} - \mu N - {\mathsf{L}}_{q}^{(3)}\Big)\bigg)\right)^{q}. \nn
\label{qw_gc_def}
\eea
A perturbative computation of the sum of $q$-weights is evolved in the following manner, first the $q$-exponential
is expressed in terms of the ordinary exponential
\bea
{\mathfrak{c}}^{(3)}(\mu,V,\beta) &=& \frac{1} {\left({\bar{Z}}_{q}^{(3)}(\mu,V,\beta)\right)^{q}}  \;
                                      \widehat{{\mathcal{D}}}_{2}(d_{\beta}) \;
                                      \exp({\mathcal{L}})\, \times
\label{qw_ps_gc}                                      \\
                                  & &  \times \, \sum_{N=0}^{\infty}
                                      \frac{1}{N!\, h^{DN}}
                                      \int {\rm d}^{DN} r  \;  {\rm d}^{DN} p \;
                                      \exp\bigg(-{\tilde{\beta}} \Big(\sum_{i}
                                      \frac{p_{i}^{2}}{2m} - \mu N\Big)\bigg), \nn
\eea
where the operator series $\widehat{{\mathcal{D}}}_{2}(d_{\beta})$ appearing in (\ref{qw_ps_gc}) is
\beq
\widehat{{\mathcal{D}}}_{2}(d_{\beta}) = 1 - (1-q) \,\Big(d_{\beta}^{(1)} + \frac{1}{2} \; d_{\beta}^{(2)}\Big)
                                         + (1-q)^{2} \;\left(d_{\beta}^{(2)} + \frac{5}{6} \; d_{\beta}^{(3)}
                                         + \frac{3}{8} \; d_{\beta}^{(4)}\right)
                                         + \cdots.
\label{qw_deriv_ser}
\eeq
In the second step the phase space integral and the summation over the number of particles $N$ is carried out in
(\ref{qw_ps_gc}) and the resultant expression is
\beq
{\mathfrak{c}}^{(3)}(\mu,V,\beta) = \frac{1} {\left({\bar{Z}}_{q}^{(3)}(\mu,V,\beta)\right)^{q}} \;\;
                                    \widehat{{\mathcal{D}}}_{2}(d_{\beta}) \; \exp({\mathcal{L}}) \;
                                    Z_{BG}({\tilde{\beta}}).
\label{qw_BGpf_rel}
\eeq
In the final step, the rhs in (\ref{qw_BGpf_rel}) is computed and the
series corresponding to the generalized partition function (\ref{gen_par_pert}) is
substituted to obtain the perturbative series corresponding to the sum of $q$-weights
\beq
\mathfrak{c} = 1 + (1-q) \ \mathfrak{P}_{1} + (1-q)^{2} \ \mathfrak{P}_{2} + \cdots,
\label{prob_pert_impl}
\eeq
where the perturbative coefficients $\mathfrak{P}_{1}$ and $\mathfrak{P}_{2}$ are
\bea
\mathfrak{P}_{1} &=& - N_{BG}({\tilde{\beta}}) \, (\a-1)\nn \\
\mathfrak{P}_{2} &=& - \frac{1}{2} \, \u \,(\u-1) + \frac{1}{2} \, (N_{BG}({\tilde{\beta}}))^{2}\, (2 \,(\a)^{3}
                     + (\a)^{2} + (2-D) \,\a)   \\
                 & & + \frac{1}{4} \, N_{BG}({\tilde{\beta}}) (2 \, (\a)^{3} + 2 \,(2 \,\u + 1) \, (\a)^{2} + 6D \, \a
                     - (2 \, \u - 1) \,(4-D) + 4 \, \u) \nn
\label{prob_impl}
\eea
The integral form of the grandcanonical heat function, the Hill energy ${\mathsf{L}}_{q}^{(3)}$ in
the third constraint picture is
\bea
{\mathsf{L}}_{q}^{(3)}(\mu,V,\beta) &=& \frac{1} {\left({\bar{Z}}_{q}^{(3)}(\mu,V,\beta)\right)^{q}}
                                        \sum_{N=0}^{\infty} \frac{1}{N!\, h^{DN}}
                                        \int {\rm d}^{DN} r  \;  {\rm d}^{DN} p  \;
                                        \Big(\sum_{i} \frac{p_{i}^{2}}{2m} - \mu N\Big) \, \times  \\
                                    & & \phantom{aaaaaaaaaaaaaaaaaaaaa} \times \, \left(\exp_{q}\bigg(-{\tilde{\beta}} \Big(\sum_{i}
                                        \frac{p_{i}^{2}}{2m} - \mu N - {\mathsf{L}}_{q}^{(3)}\Big)\bigg)\right)^{q}. \nn
\label{HE_gc_def}
\eea
Using a perturbative procedure akin to the one employed in the sum of $q$-weights the Hill energy in terms of the
derivative series (\ref{qw_deriv_ser}) is
\beq
{\mathsf{L}}_{q}^{(3)}(\mu,V,\beta) = \frac{1} {\left({\bar{Z}}_{q}^{(3)}(\mu,V,\beta)\right)^{q}} \;
                             \widehat{{\mathcal{D}}}_{2}(d_{\beta}) \left(\frac{D}{2 \, {\tilde{\beta}}} - \mu\right) \;
                              \exp({\mathcal{L}})\; N_{BG}({\tilde{\beta}}) \;
                             Z_{BG}({\tilde{\beta}}).
\label{HE_gc_Deriv}
\eeq
Calculating the derivatives in the above expression and substituting the Boltzmann-Gibbs partition function,
the perturbative series of the Hill energy upto $(1-q)^{2}$ order is
\beq
\mathsf{L}_{q}^{(3)} = \mathsf{L}_{BG}(\tilde{\beta}) (1 + (1-q)\ \mathsf{L}_{1} + (1-q)^{2} \ \mathsf{L}_{2} + \cdots).
\label{HE_3c_ps}
\eeq
where $\mathsf{L}_{BG}(\beta) = \left(\frac{D}{2 \, \beta} - \mu\right) \; N_{BG}(\beta)$ and the coefficients of
perturbation are listed below
\bea
\mathsf{L}_{1} &=& \alpha(\tilde{\beta}) + 2\, \mathcal{L} + (1 + \alpha(\tilde{\beta}))\, N_{BG}(\tilde{\beta}), \nn \\
\mathsf{L}_{2} &=& - \frac{1}{8} \; \big( 4\, (\alpha(\tilde{\beta}))^{3} - 4 \, (2 + D - 4\, (1 - D \mu)\; \tilde{\beta}
                   - 4 (\tilde{\beta} \mu)^{2})\; \mathcal{L} + 4\,(3 + 4 \tilde{\beta} \mu ) \; \mathcal{L}^{2}
                   + 8 \nn \\
               & & + 2 D \;  {\tilde{\beta}} \mu - 8\, \tilde{\beta}^{2}\big) - \frac{1}{2}\; N_{BG}(\tilde{\beta})\;
                     \big(2 - D + 4\, D^{2} \,
                     \tilde{\beta} \mu - 8\; (\tilde{\beta} \mu)^{2} - 6 \, (\alpha(\tilde{\beta}))^{3} + \big(3D  \nn \\
               & & - 2 D^{2} + 8\, (1 - D) \; \tilde{\beta} \mu  - 8 \, (\tilde{\beta} \mu)^{2}\big)\; {\mathcal{L}}\big)
                   - \frac{1}{8}\; (N_{BG}(\tilde{\beta}))^{2} \;  \big(3 D^{2} - 16 D \; \tilde{\beta} + 20 \tilde{\beta}^{2} \nn \\
               & & - 16 \; (\alpha(\tilde{\beta}))^{3} \big).
\label{HE_3c_terms}
\eea
From the perturbative solutions we notice that the generalized ensemble (\ref{gen_par_pert}),
the sum of $q$- weights (\ref{prob_pert_impl}) and the Hill energy (\ref{HE_3c_ps}) are coupled equations.
The explicit expression corresponding to these quantities are obtained by employing a recursive procedure
to solve these equations.  The important characteristic feature that all these equations are uncoupled at
$q=1$ enables us to obtain a solution. Through the use of the recursive procedure, an explicit form of the
generalized partition function is obtained up to second order in $(1-q)$ and reads:
\beq
Z_{q}^{(3)}(\beta) = Z_{BG}(\beta)(1 + (1-q) \ \mathcal{Z}_{1}(\beta) + (1-q)^{2} \ \mathcal{Z}_{2}(\beta) + \cdots).
\label{gen_gce_ps}
\eeq
where the coeffecients of perturbation are
\bea
\mathcal{Z}_{1}(\beta) &=& - \frac{1}{16} \big(4D + 4D^{2} + D^{3} - (4D - 2D^{2}) \beta \mu + (8 - 4D) \, (\beta \mu)^{2}
                           + 8 (\beta \mu)^{3}\big) N_{BG}(\beta)\nn \\
                       & & + \frac{1}{8} \big(4D + 4D^{2} + D^{3} - (8D + 4D^{2}) \beta \mu + 8(1+D)\, (\beta \mu)^{2}
                           - 8 (\beta \mu)^{3} \big) (N_{BG}(\beta))^{2}, \nn \\
\mathcal{Z}_{2}(\beta) &=& + \frac{1}{768} \big(32D + 88D^{2} + 28D^{3} + 2D^{4} + 3 D^{5} - (32 D + 168 D^{2} + 40 D^{3} \nn \\
                       & & - 18 D^{4}) \, \beta \mu + (96D + 96D^{2} - 24D^{3}) \, (\beta \mu)^{2} - (512 - 416 D + 240 D^{3}) \,
                             (\beta \mu)^{3} \nn \\
                       & & - (544 - 432D)\, (\beta \mu)^{4} - 96\, (\beta \mu)^{5}\big)\, N_{BG}(\beta) + \frac{1}{512}
                             \big(176D^{2} - 32D^{4} + 8D^{5} + D^{6} \nn \\
                       & & - (160 D^{2} - 432 D^{3} - 8 D^{4} - 4 D^{5})\, \beta \mu + (576D - 656D^{2} - 160D^{3}  \nn \\
                       & & + 12 D^{4})\, (\beta \mu)^{2} - (1024 + 640D - 704D^{2} + 32 D^{3}) \, (\beta \mu)^{3} - (704
                           + 1024 D \nn \\
                       & & - 48 D^{2})\, (\beta \mu)^{4} + (384 - 64D)\, (\beta \mu)^{5} + 64\, (\beta \mu)^{6}\big) \,
                             (N_{BG}(\beta))^{2} - \frac{1}{128} \, \big(32D \nn \\
                       & & + 48D^{2} + 16D^{3} + 2D^{5} + D^{6} - (32D - 32 D^{2} - 96 D^{3} - 24 D^{4} + 6 D^{5})\, \beta \mu
                             \nn \\
                       & & + (192 - 64D - 208D^{2} - 152D^{3} - 20 D^{4}) (\beta \mu)^{2} - (320 - 128 D - 368 D^{2} \nn \\
                       & & + 48 D^{3})\, (\beta \mu)^{3}  - (128 + 448 D - 80 D^{2}) (\beta \mu)^{4} + (192 - 96 D)\,
                             (\beta \mu)^{5}\nn \\
                       & & - 64\, (\beta \mu)^{6}\big)\, (N_{BG}(\beta))^{3} + \frac{1}{128}\, \big(16 D^{2} + 32 D^{3}
                           + 24 D^{4} + 8 D^{5} + D^{6} - (64 D^{2} \nn \\
                       & & + 96 D^{3} + 48 D^{4} + 8 D^{5})\, \beta \mu + (64D + 192 D^{2} + 144 D^{3} + 32 D^{4})\, (\beta \mu)^{2}
                           - (192D \nn \\
                       & & + 256 D^{2} + 80 D^{3})\, (\beta \mu)^{3} + (64 + 256 D + 128 D^{2})\, (\beta \mu)^{4} - 128(1+D)\,
                             (\beta \mu)^{5} \nn \\
                       & & + 64\, (\beta \mu)^{6}\big) \, (N_{BG}(\beta))^{4}.
\eea
A similar application of the recursive procedure enables us to calculate the explicit expression of the
Hill energy upto second order in $(1-q)$.  The perturbative series corresponding to the Hill energy is
\beq
{\mathsf{L}}_{q}^{(3)}(\beta) =   {\mathsf{L}}_{BG}(\beta) \left(1 + (1-q) \ {\mathbf{L}}_{1} (\beta)
                                + (1-q)^{2} \ \mathbf{L}_{2} (\beta) + \cdots\right),
\label{He_GC_expl}
\eeq
and the coefficients of perturbation are listed below
\bea
\mathbf{L}_{1}(\beta) &=& - \bigg(\frac{1}{16}\; \big(D^{3} + 2D^{2} + 12D \; \beta \mu + 4D \; \beta \mu \; \alpha(\beta)
                          + 16\; (\beta \mu)^{2} + 8 (\beta \mu)^{3}\big) + \frac{1}{8} \; \big(4D  \nn \\
                      & & + 2D^{2} + D^{3} + 8D \; \beta \mu \; \alpha(\beta) + 8 (\alpha(\beta))^{2}
                          - 8 (\beta \mu)^{3}\big)\; N_{BG}(\beta)\bigg) \; U_{BG}(\beta), \nn \\
\mathbf{L}_{2}(\beta) &=& \bigg(\frac{1}{768} \big(16 D^{2} - 12 D^{3} - 4 D^{4} + 3 D^{5} - (32D + 72D^{2} - 8 D^{3} - 18D^{4})
                          \; \beta \mu  \nn \\
                      & & + (192D - 48 D^{2} -24 D^{3}) \; (\beta \mu)^{2} - (768 - 608D - 240D^{2})\; (\beta \mu)^{3} - (640 \nn \\
                      & & - 432 D)\; (\beta \mu)^{4} - 96\; (\beta \mu)^{5}\big) + \frac{1}{16}\;  \big(4D^{2} - 2 D^{3}
                          - (8D + 10 D^{2} - 13D^{3}) \; \beta \mu  \nn \\
                      & & + (32D - 18 D^{2} - 5 D^{3}) \; (\beta \mu)^{2} + 28D \; (\beta \mu)^{4} + 8\; (\beta \mu)^{5}\big) \;
                            N_{BG}(\beta) + \frac{1}{64} \; \big(8D^{2}  \nn \\
                      & & + 20 D^{3} + 14D^{4} + 3D^{5} - (16D + 56D^{2} + 76D^{3} + 26D^{4})\; \beta \mu
                          - (64 - 80D  \nn \\
                      & & - 128D^{2} - 100D^{3})\; (\beta \mu)^{2} + (96 - 80D - 208D^{2})\; (\beta \mu)^{3} + (64 + 224D)\;
                            (\beta \mu)^{4} \nn \\
                      & & - 96 (\beta \mu)^{5}\big)\; (N_{BG}(\beta))^{2}\bigg) \; U_{BG}(\beta).
\label{He_expl_pt}
\eea
The knowledge of the explicit form of the generalized partition function (\ref{gen_gce_ps}) and the
Hill energy (\ref{He_GC_expl}) enables us to compute the average number of particles and the internal energy.
First the entropy of the system is calculated from the generalized partition function (\ref{gen_gce_ps}) through
the use of (\ref{entropy_3c}).  In the next step the explicit form of the Hill energy (\ref{He_GC_expl}) is used in
conjunction with the entropic expression in the relation $\Phi_{q}^{(3)} = {\mathsf{L}}_{q}^{(3)} - T S$ to
obtain the free energy. From the free energy the average number of particles can be found through the use of a
specific form of (\ref{Vol_hf_Gfe}) i.e., $N_{q}^{(3)} = - \frac{\partial \Phi_{q}^{(3)}} {\partial \mu}$.
The explicit expression corresponding to the average number of particles in the third constraint $N_{q}^{(3)}$ is
\beq
N_{q}^{(3)} = \zn + (1-q) \ {\mathfrak{N}}_{1}(\beta) + (1-q)^{2} \ {\mathfrak{N}}_{2}(\beta) + \cdots,
\label{nop_pert_expl}
\eeq
where the terms ${\mathfrak{N}}_{1}$ and ${\mathfrak{N}}_{2}$ are
\bea
{\mathfrak{N}}_{1}(\beta) &=& -\frac{1}{4} \; \zn \; \big((2 \; (\b)^{2} + 4 \;\b-D) - 4 \, \zn \;((\b)^{2} - \b)\big) \nn \\
{\mathfrak{N}}_{2}(\beta) &=& \frac{1}{96} \; \zn \; (12 \; (\b)^{4} + 80 \; (\b)^{3} + 12(8-7D)\bb^{2}
                              - 6D^{2}(2-D)^{2}\; \beta \mu \nn\\
                          & & + D(1-D)(4+D)) - \frac{1}{8} \;(\zn)^{2}\; (4\; (\b)^{4} + 8 \; (\b)^{3} + 8D (\b)^{2}\nn \\
                          & & - 16\; \beta \mu \; \b - D(4+D)) + \frac{1}{16}\;(\zn)^{3} \; (24 (\b)^{4}
                              - 16\; \beta \mu (1+\beta \mu)\b \nn \\
                          & & - 8(1-8D)\bb^{2} + 8(2-3D^{2})\beta \mu - D(4-2D-5D^{2})).
\label{nop_expl}
\eea
The calculated results corresponding to the Hill energy (\ref{He_GC_expl}) and the average number of
particles (\ref{nop_pert_expl}) is used in the  relation $\mathsf{L}_{q}^{(3)} = U_{q}^{(3)} - \mu N_{q}^{(3)}$
to find the internal energy. The perturbative series corresponding to the internal energy is
\beq
U_{q}^{(3)}(\beta) = U_{BG}(\beta) + (1-q) \ {\mathcal{U}}_{1} + (1-q)^{2} \ {\mathcal{U}}_{2} + \cdots,
\label{int_pert_expl}
\eeq
where the perturbative terms are listed below
\bea
{\mathcal{U}}_{1}(\beta) &=& - \frac{1}{2}\, \big(\bb^{2} - 2 \,\beta \mu + D(2+D) + 2 \,(D \,\b + 2 \; \beta \mu \nn \\
                         & & - 2 \, (1+D)) \, \zn\big) \, U_{BG}(\beta),\nn \\
{\mathcal{U}}_{2}(\beta) &=&  \frac{1}{96} \, \Big((12 \, (\b)^{4} + 48 \, \bb^{4} - 16 \, \bb^{3} + 36D(1-D) \,\bb^{2}
                             + 12D(2-D^{2}) \,\beta \mu \nn\\
                         & & + D(4-3D-D^{2})) - 4 \, \zn \, \big(24 \, \bb^{3} + 42D \,\bb^{2} - 88D^{2} \, \beta \mu\nn\\
                         & & - 6D(1-D)(2+D) - \b \,(36 \, \bb^{3} + 12D \, \bb^{2} + 3(16+D^{2})\, \beta \mu)\big)\nn \\
                         & & + 3\, (\zn)^{2} \,(32 \, (\b)^{4} + 4 (8+D^{2}) \, \b  \, \beta \mu
                             + 16 \, \bb^{3} + 32 (5+2D)\, \bb^{2} \nn \\
                         & & - (32 + 32D - 56D^{2} - 2D^{3}) \beta \mu + D \,(8 + 20 \,D + 14 D^{2}+
                               D^{3})\Big)\, U_{BG}(\beta).
\label{int_expl}
\eea
The average number and the internal energy can also be calculated in a different manner wherein the basic definition
of these quantities in (\ref{ge_ie_3c},\ref{ge_av_3c}) is used to find the implicit expressions.  The implicit
expressions are then solved to arrive at the final explicit forms.  The results obtained through this method
agrees precisely with those obtained with those in (\ref{nop_pert_expl}) and (\ref{int_pert_expl}).  This is
a nontrivial consistency check on the results obtained for the physical quantities.

\par

The entire calculation has been carried out directly in the third constraint picture and consistently upto second order
in the expansion parameter $(1-q)$. However the process is not limited only to the second order and it can be
extended to any arbitrary order in the expansion parameter.
\subsection{Extreme relativistic classical ideal gas}
The perturbative study carried out for the nonrelativistic classical ideal gas can also be extended to the
system of extreme relativistic classical ideal gas.  Since the calculational procedure is same as
in the nonrelativistic case detailed in the previous section, we present only the final expressions
corresponding to the internal energy and the average number of particles. The average number of particles is
\beq
\mathtt{N}_{q}^{(3)}(\beta) = \mathtt{N}_{BG}(\beta) + (1-q)\ \mathtt{N}_{1}(\beta) + (1-q)^{2}\ \mathtt{N}_{2}(\beta) + \cdots,
\label{avg_gc_Urg}
\eeq
where the term $\mathtt{N}_{BG} = \exp( \beta \mu) \,V \, \Delta\, \beta^{-D} $ and the coefficients of perturbation are
\bea
\mathtt{N}_{1}(\beta) &=& - \frac{1}{2}\; \mathtt{N}_{BG}(\beta)\; \big((\beta \mu)^{2} + 2 (1+D)\; \beta \mu - D(1-D)\big)
                          + (\mathtt{N}_{BG}(\beta))^{2} \; \big((\beta \mu)^{2}  \nn \\
                      & & - (2+D)\; \beta \mu + 3D - 1\big), \nn \\
\mathtt{N}_{2}(\beta) &=& - \frac{1}{24}\; \mathtt{N}_{BG}(\beta)\; \big(3 \; (\beta \mu)^{4} + (20 - 12D)\; (\beta \mu)^{3}
                          + (24 - 42D) (\beta \mu)^{2} \nn \\
                      & & - 12D(1 - 2D - D^{2})\; \beta \mu + 2D - 3D^{2} - 2D^{3} + 3D^{4} - \frac{1}{2}\;
                           (\mathtt{N}_{BG}(\beta))^{2} \; \big((\beta \mu)^{4} \nn \\
                      & & - 2D\; (\beta \mu)^{3} - (8 - 10D)\; (\beta \mu)^{2} -  (6 - 12D + 16 D^{2} - 2D^{3}) \; \beta \mu \nn\\
                      & & - D^{2}(1 - 6D + D^{2})\big) + \frac{1}{2}\; (\mathtt{N}_{BG}(\beta))^{3} \; \big(3\; (\beta \mu)^{4}
                          - (9 - 6D)\; (\beta \mu)^{3} \\
                      & & - (2 - 28 D - 3 D^{2})\; (\beta \mu)^{2} + (22 - 20D - 19 D^{2})\; \beta \mu + 2 - 11D + 20 D^{2}
                          + D^{3}\big).\nn
\eea
The final expression for the internal energy is
\beq
\mathtt{U}_{q}^{(3)}(\beta) = \mathtt{U}_{BG}(\beta) + (1-q)\ \mathtt{U}_{1}(\beta) + (1-q)^{2}\ \mathtt{U}_{2}(\beta) + \cdots,
\label{IE_gc_Urg}
\eeq
and the coefficients of perturbation are listed below
\bea
\mathtt{U}_{1}(\beta) &=&  -((\beta \mu)^{2} - 2 D \; \beta \mu + D(1+D))\mathtt{U}_{BG}(\beta) +
                           \mathtt{N}_{BG}(\beta) ((\beta \mu)^{2} - (3+D) \beta \mu \nn \\
                      & &  + (1+3D)) \mathtt{U}_{BG}(\beta), \nn \\
\mathtt{U}_{2}(\beta) &=&  \bigg(\frac{1}{24}\, \big(3\, (\beta \mu)^{4} + (8-12D)\, (\beta \mu)^{3}  - 6D (1+3D)\,
                          (\beta \mu)^{2} - 12D^{2}(D+1) \, \beta \mu      \nn\\
                      & &  + 2D + 6 D^{2} + 10 D^{3} + 3D^{4}\big) - \frac{1}{2} \, \mathtt{N}_{BG}(\beta)
                           \big((\beta \mu)^{4} - 2 (1+D) \, (\beta \mu)^{3} \nn \\
                      & &  - (4 - 2D) \, (\beta \mu)^{2} - (4D +14 D^{2} - 2D^{3})\, \beta \mu
                           + 2D + 7 D^{2} - 4 D^{3} + 2 D^{4}\big)   \nn \\
                      & &  + \frac{1}{2}\, (\mathtt{N}_{BG}(\beta))^{2} \, \big(3 \, (\beta \mu)^{4}
                           + (13 - 6D) \, (\beta \mu)^{3} + (13 - 32D + 3D^{2}) \, (\beta \mu)^{2} \nn \\
                      & &  -(1 + 40D + 19 D^{2})\, \beta \mu + 1 + 9D + 21D^{2} + D^{2} \big)\bigg)\, \mathtt{U}_{BG}(\beta).
\eea
and the Boltzmann-Gibbs factor is $\mathtt{U}_{BG}= \frac{D} {\beta} \; \mathtt{N}_{BG}$.
The Boltzmann-Gibbs values for both the quantities are recovered in the $q \rightarrow 1$ limit.

\par

The thermodynamic results corresponding to a one dimensional model of an interacting gas known as Tonks gas described by the
Hamiltonian (\ref{ham_tg}) can be read from the results of the classical nonrelativistic ideal gas.  For this purpose we employ
a special kind of grandcanonical ensemble where the free length or the total length available to the molecules is used instead
of volume.  Also the number of dimensions is set to be $1$ and the factor $\mathcal{M}_{1}$ the one dimensional analog of
$\mathcal{M}$ is used.
%
%
%
%
\setcounter{equation}{0}
\section{Generalized ensemble}
\label{GE}
The generalized ensemble describes a completely open system which can
exchange the internal energy, the volume and the number of particles.  The
heat function is the ${\mathsf{R}}$ function described by ${\mathsf{R}} = U + PV - \mu N$
and the generalized free energy is the function
${\mathcal{E}} = U - T S + PV - \mu N$.
\subsection{Nonrelativistic ideal gas}
The classical nonrelativistic ideal gas is analyzed in the generalized ensemble
picture. A perturbative process based on the idea of disentangling of
$q$-exponential [\cite{CQ}] is used. Though the thermodynamic quantities are evaluated uniformly
up to $(1-q)^{2}$ order, the results are presented only up to the first order in
$(1-q)$ since some of the quantities are too cumbrous to be displayed up to second
order. The generalized partition function in the third constraint picture is
\beq
{\bar{\Upsilon}}_{q}^{(3)}(\mu,V,\beta) = \sum_{N=0}^{\infty} \frac{1}{N!\, h^{DN}}\int
                                          {\rm d}^{DN} r  \;  {\rm d}^{DN} p \;\;
                                          \exp_{q} \left(-{\hat{\beta}}\left(\sum_{i}
                                          \frac{p_{i}^{2}}{2m} + P V - \mu N -
                                          {\mathsf{R}}_{q}^{(3)}\right)\right),
\label{gpf_GE_def}
\eeq
where the factor ${\hat{\beta}} = \frac{\beta}{{\mathfrak{c}}^{(3)}}$ and
${\mathfrak{c}}^{(3)}$ is the third constraint sum of $q$-weights in the generalized ensemble.
The heat function ${\mathsf{R}}_{q}^{(3)} = U_{q}^{(3)} + P V_{q}^{(3)} - T S - \mu N_{q}^{(3)}$
where $U_{q}^{(3)},V_{q}^{(3)},$ and $N_{q}^{(3)}$ are the internal energy, average
volume and average number of particles respectively in the third constraint. The quantities
$P$ and $\mu$ denote the pressure and the chemical potential of the system.

In (\ref{gpf_GE_def}) the phase space integrals can be carried out after the due substitution of the perturbative expansion of the $q$-exponential in terms of the ordinary exponential. Since the volume states are very close to each other, we approximate the
volume summation by an integral.  Employing the correct counting of volume states advocated in [\cite{DC1},\cite{DC2}], we
carry out the volume integral.  Finally we assume that the summation over the number of particles is convergent atleast, in the
neighbourhood of $q=1$ and, thus arrive at the final perturbative series of the generalized partition function
\beq
{\bar{\Upsilon}}_{q}^{(3)}(\mu,P,\beta) =  \widehat{{\mathcal{D}}}_{1}(d_{\beta}) \;
                                           \exp\left({\mathcal{R}}\right) \;
                                           \Upsilon_{BG}({\hat{\beta}}),
\label{GE_BGpf_rel}
\eeq
where ${\mathcal{R}} = {\hat{\beta}} \; {\mathsf{R}}_{q}^{(3)}$ and $\widehat{{\mathcal{D}}}_{1}(d_{\beta})$ is
the derivative series defined in (\ref{gpf_deriv_ser}) .  The partition function and the average number of particles in the Boltzmann-Gibbs
framework is
\beq
\Upsilon_{BG}(\beta) = \frac{1}{1- \om}, \qquad N_{BG}(\beta) = \frac{\om}{1-\om},  \qquad
\om = \exp(\beta \mu) \;  \mathcal{M} \; \frac{1}{P}\; \left(\frac{1}{\beta}\right)^{\frac{D}{2} + 1},
\label{BGq_def}
\eeq
where the factor $\om$ is considered to be less than $1$ so as to maintain the positivity of the partition function.
From (\ref{GE_BGpf_rel}) the perturbative series corresponding to the third constraint generalized partition function
upto $(1-q)$ order is computed
\bea
{\bar{\Upsilon}}_{q}^{(3)}(\mu,P,\beta) &=& \exp\left({\mathcal{R}}\right) \, \Upsilon_{BG}({\hat{\beta}}) \,
                                           \Big(1 + (1-q) \ \frac{1}{4} \, \Big((2\, (\ag)^{2}
                                           + 4 R\, {\hat{\beta}} \, \ag  - 4 \, (R + \mu)\, {\hat{\beta}} \nn \\
                                       & & + (4 + 3D)) N_{BG}(\hat{\beta}) + (2\, (\ag)^{2} - 4  \ag + 2) (N_{BG}(\hat{\beta}))^{2}\Big) + \cdots \Big).
\label{GE_gpf_ps}
\eea
Applying a similar perturbative technique, the sum of $q$-weights in the generalized ensemble picture reads:
\beq
{\mathfrak{c}}^{(3)}(\mu,P,\beta) = \frac{1} {\left({\bar{\Upsilon}}_{q}^{(3)}(\mu,P,\beta)\right)^{q}} \;\;
                                    \widehat{{\mathcal{D}}}_{2}(d_{\beta}) \; \exp({\mathcal{R}}) \;
                                    \Upsilon_{BG}({\hat{\beta}}),
\label{qw_BGpf_GErel}
\eeq
where the derivative series $\widehat{{\mathcal{D}}}_{2}(d_{\beta})$ is defined in (\ref{qw_deriv_ser}).
Employing the derivative series, the perturbative expression of the sum of $q$-weights upto first order in
$(1-q)$ is
\beq
{\mathfrak{c}}^{(3)}({\hat{\beta}}) = 1 + (1-q)\ \Big(\big((\ag + 1) - \Upsilon_{BG}(\hat{\beta})\big)\; \omega(\hat{\beta}) -
                                      \Upsilon_{BG}(\hat{\beta})\Big) \; \ln \Upsilon_{BG}(\hat{\beta}) + \cdots.
\label{qw_GE_ps}
\eeq
Adopting a procedure identical to the one used for the generalized partition function and the sum of $q$-weights,
the ${\mathsf{R}}_{q}^{(3)}$ heat function can be written as:
\beq
{\mathsf{R}}_{q}^{(3)}(\mu,P,\beta) = \frac{1} {\left({\bar{\Upsilon}}_{q}^{(3)}(\mu,P,\beta)\right)^{q}} \;
                                      \widehat{{\mathcal{D}}}_{2}(d_{\beta}) \left(\left(\frac{D}{2}+1\right) \,
                                      \frac{1} {\hat{\beta}} - \mu\right) \; \exp({\mathcal{R}})\; N_{BG}({\hat{\beta}}) \;
                                      \Upsilon_{BG}({\hat{\beta}}).
\label{HE_GE_Deriv}
\eeq
The perturbative series of the ${\mathsf{R}}_{q}^{(3)}$ upto $(1-q)$ order found from (\ref{HE_GE_Deriv}) is
\beq
{\mathsf{R}}_{q}^{(3)}(\mu,P,\beta) = \left(\left(\frac{D}{2}+1\right) \,
                                      \frac{1} {\hat{\beta}} - \mu\right)\,  N_{BG}(\tb) + (1-q)\ \left(\mathsf{R}_{1}
                                     + \mathsf{R}_{2} + \mathsf{R}_{3}\right) + \cdots.
\label{Rhf_GE_ps}
\eeq
The perturbative coefficients of $(1-q)$ term in (\ref{Rhf_GE_ps}) is
\bea
\mathsf{R}_{1} &=&  \bigg(R^{2} \, \tb (1-\alpha(\hat{\beta})) +  \frac{1}{16} \, \Big(\big(4(D+2)(D+4)\, P^{-1} \tb^{-2} -(D+2)((D+4)(D+6) \nn \\
               & & + 8\, (R + 2\,\mu)P^{-1}) \tb^{-1} + (R + \mu)\, \mu P^{-1} + 3\, \mu - 4\,((D+2)\, (R + 4\, \mu)R   \nn \\
               & & - (2\, R^{2} - 6\, \mu - 3D\, \mu)\mu)+ 6D (D+6) \, \mu + 8 \, (R + \mu)\, (\tb \mu)^{2}\big)\nn \\
               & & + \omt\, \big(2 \,(D^{3} + 40D^{2} + 32D + 32 + 2(D+2)\, R^{2} + 4(D+2) \, (R + 2\, \mu)) \, P^{-1}\tb^{-1} \nn \\
               & & - 4 \,(D+2)(D+4)\, P^{-1}\tb^{-2} - 4(2(D^{2} + 5D + 3)\, R + (D+2)(3D+10)\, \mu  \nn \\
               & & + 4\, (R+\mu)\, \mu P^{-1}) + 8((2D+4)\, R + 3D \, \mu)\, \tb \mu -((R + 2\mu)^{2} \nn \\
               & & - 2\, \mu^{2})\, (\tb \mu)^{2}\big)\Big)\, \up \bigg)\,  N_{BG}(\tb),\nn \\
\mathsf{R}_{2} &=&  \frac{1}{8}\,  \Big((D+2)^{3} \, \tb^{-1} + (D+2)((D+2)(3D+8) + 12 \, \mu^{2}) - 2\, (12 \, \mu^{3}P^{2} + 4 \, \mu^{3}P^{-1}\nn \\
               & & + 4(9D^{2}+34D+32)\, \mu P^{2} + 2(D+2)^{2}\, R P^{2})\, \tb^{2} + 4\, ((9D+14)\mu^{2} \nn \\
               & & + 2(D+2) \, R \mu P^{2})\tb^{3} - 12(D+2)^{2} \, \mu) - ((3D^{3}+20D^{2}+44D+32)\, P^{2}\tb \nn \\
               & & - 2 \, ((9D^{2}+34D+32)\, \mu + 2(D+2)^{2}\, R)\, \tb P^{2}  - 4\, (4(D+2)R  \nn \\
               & & + (9D+14) \, \mu) \, P^{2}\tb^{3}\mu + (2R + 3\mu)\, \tb^{5}\mu^{3}P^{2})\Big)\up \, (N_{BG}(\tb))^{2}, \nn \\
\mathsf{R}_{3} &=&   \frac{1}{8}\, \Big(3 \, \omt((D+2)^{3}\, P^{3}\tb^{2} - 6(D+2)^{2}\, \mu P^{3}\tb^{3} + 12(D+2)\,  \mu^{2}\tb^{4}P^{3}
                   - 8 \, \mu^{3}\tb^{5}P^{3})\nn \\
               & & - 3((D+2)^{3}\,  \tb^{2}P^{3} - 6(D+2)^{2}\,  \mu \tb^{3}P^{3} + 12 (D+2)\, \tb^{4}\mu^{2}P^{3} \nn \\
               & & - 8 \, \tb^{5}\mu^{3}P^{3})\Big)\, \up \, (N_{BG}(\tb))^{3}.
\label{Rfun_imp}
\eea
The generalized partition function (\ref{GE_gpf_ps}), the sum of $q$-weights (\ref{qw_GE_ps}), and
the $\mathsf{R}_{q}^{(3)}$ heat function (\ref{Rhf_GE_ps}), form a set of coupled implicit perturbative expansions.
The fact that these equations are uncoupled at the lowest perturbative order allows us to obtain an explicit form of
these quantities through the use of a recursive procedure. The final explicit form of the generalized partition function
computed upto first order in the expansion parameter $(1-q)$ is
\beq
\bar{\Upsilon}_{q}^{(3)}(\beta) = \Upsilon_{BG}(\beta) \; \exp(\beta R) \; \left(1 + (1-q) (\Upsilon_{1}(\beta)
                                  + \Upsilon_{2}(\beta) + \Upsilon_{3}(\beta) + \Upsilon_{4}(\beta)) + \cdots \right),
\label{gpf_GE_Exp}
\eeq
and the coefficients of perturbation are listed below:
\bea
\Upsilon_{1}(\beta) &=& - \frac{1}{16} \Big(4 \, (2 \, (\alpha(\beta))^{2} - 4 \, \alpha(\beta) + D)  -  \big((2+D)
                          \, \frac{1}{\beta P} \big((2+D)(4 + D - \beta \mu)\nn \\
                    & & - 4 (\beta \mu)^{2}\big) - 4 (4 + 6D + D^{2} + 2(2+D)\, \beta \mu) \ln \Upsilon_{BG}(\beta)\big)\,
                          \Upsilon_{BG}(\beta)\nn \\
                    & & + 2\, \big(D (2+D)\, \beta \mu - 4 \,(\beta \mu)^{3} - 4\, (2(2+D) - 2(1-D)\, \beta \mu  \nn \\
                    & & - (\beta \mu)^{2})  \ln \Upsilon_{BG}(\beta)\big)(\Upsilon_{BG}(\beta))^{2}
                        + 16 \, \frac{1}{(\beta P)^{2}}(1 + \alpha(\beta)) (\Upsilon_{BG}(\beta))^{3}\Big)\, N_{BG}(\beta), \nn \\
\Upsilon_{2}(\beta) &=& - \frac{1}{8} \Big((6 + 3D - (8 + D)\, \beta \mu - (\beta \mu)^{2} - 2 \, (1 + D - \beta \mu)
                          \ln \Upsilon_{BG}(\beta)) \nn \\
                    & & + (16 + 8D - 2 D^{2} - 16 \beta P - (8 - 10D + D^{2} + 8 \beta P - 4D\, \beta P) \beta \mu
                           \nn \\
                    & & - (8 + 4D + 8 \beta P)\, (\beta \mu)^{2} + 12\, (\beta \mu)^{3} + 4 (6 + 3D - 2 (\beta P)^{-1}
                        - (4 - D)\, \beta \mu \nn \\
                    & & + (\beta \mu)^{2})\,  \ln\Upsilon_{BG}(\beta))\, \Upsilon_{BG}(\beta) - 4\, ((\beta P)^{-2}
                        - 4\, (\beta P)^{-1} \alpha(\beta) \nn \\
                    & & - (\alpha(\beta))^{2})\, (\Upsilon_{BG}(\beta))^{2} \Big)\, (N_{BG}(\beta))^{2}, \nn \\
\Upsilon_{3}(\beta) &=& - \frac{(\Upsilon_{BG}(\beta))^{3}}{\beta P} \, \ln \Upsilon_{BG}(\beta) - \frac{1}{4} \,
                          \big(8 + 6D + D^{2} - (16 + 4D + 4 \beta P)\, \beta \mu  \nn \\
                    & & + 12\, (\beta \mu)^{2} + 4\, (1 - \alpha(\beta))\,  \ln \Upsilon_{BG}(\beta) + 4 (\alpha(\beta))^{2}
                        + (\beta P)^{-1} \alpha(\beta)\big)\, (N_{BG}(\beta))^{3}, \nn \\
\Upsilon_{4}(\beta) &=& - \frac{1}{2}\, (\alpha(\beta))^{2}\, (N_{BG}(\beta))^{4}.
\label{gpf_GE_ps}
\eea
Proceeding similarly the final explicit form of the heat function ${\mathsf{R}}_{q}^{(3)}$ is obtained upto $(1-q)$ order
\beq
{\mathsf{R}}_{q}^{(3)} =  \left(\left(\frac{D}{2}+1\right) \,\frac{1} {\beta} - \mu\right)
                         + (1-q) \ \left(\mathtt{R}_{1} + \mathtt{R}_{2} + \mathtt{R}_{3}\right) + \cdots,
\label{Rhf_GE_Exp}
\eeq
where the perturbative coefficients may be listed as
\bea
\mathtt{R}_{1} &=& \frac{1}{4} \, \Big((D+2)(D+4)\bt^{-1}-(D+2)\mu-\bt\mu^{2}\Big)\, (\upn)^{2} \, \ln \upn \, N_{BG}(\beta), \nn \\
\mathtt{R}_{2} &=& - \frac{1}{16} \,\bigg( \Big( \, (D+2)(3D+10)P - (D+2)(3\, \mu - (D^{2}-10D-40)\, P)\, P\bt  \nn \\
               & & + 2\, (24\, \mu - (D+2)(3D-16)\, P)\, \mu P \bt^{2} + 4(3D-20)\, \mu^{2}P^{2}\bt^{3} - 8\, \mu^{3}P^{2}\bt^{4}\nn \\
               & & + 2\, (8\, \mu - 3(D+2)(D+4)\, P)\, \bmu + 12(D+2)\, P\bb^{2} - 8\, P\bb^{3}\nn \\
               & & - (D+2)(16\, \mu + (D+4)(D+6)\, P)\Big)\, \upn - 2 \Big((D+2)^{2}(D+4)\bt^{-1} \nn \\
               & & - 2 (D+2)(3D+8)\, \mu^{2} + 12\, (D+2)\, \bt\mu^{2} -8 \, \bt^{2}\mu^{3}  + 2\, ((D+2)^{2}\bt^{-1}\nn \\
               & & - 4(D+2)\, \mu + 4\, \bt \mu^{2})\ln \upn \Big) \, (\upn)^{2}\bigg)\, (N_{BG}(\beta))^{2}, \nn \\
\mathtt{R}_{3} &=&  \frac{1}{8}\, \Big(2\, (D+2)((9D+10)\mu+2(D^{2}+3D+1)P)P^{2}\bt^{2} - D(D+2)(3D+8)\, P^{2}\bt \nn \\
               & & - 4\, ((9D+14) \, \mu + (D+2)(3D+4)\, P^{2})\, \mu \bt^{3} + 8\,(3\, \mu + 2(3D+5)\, P)\, \mu^{2}P^{2}\bt^{4} \nn \\
               & & - 2\,  P^{3}\mu^{3}\bt^{5}\Big)\, (N_{BG}(\beta))^{3}.
\label{Rhf_Exp_ps}
\eea
To compute the other thermodynamic quantities like the internal energy, average volume and the average number of particles
we need to calculate the free energy via the expression ${\mathcal{E}}_{q}^{(3)} = U_{q}^{(3)} + P V_{q}^{(3)}
- T S - \mu N_{q}^{(3)} = \mathsf{R}_{q}^{(3)} - T S$ .  The entropy which can be read from the generalized partition function
(\ref{gpf_GE_Exp}) using the relation (\ref{entropy_3c}), and the perturbative expansion of the heat function
$\mathsf{R}_{q}^{(3)}$ are used in this process.  From the free energy, the average volume can be found through the use
of the thermodynamic relation $V_{q}^{(3)} = \frac{\partial {\mathcal{E}}_{q}^{(3)}}{\partial P}$.  The perturbative
expansion of the average volume upto $(1-q)$ order is
\bea
V_{q}^{(3)} &=&  V_{BG}(\beta) + (1-q)\ \frac{1}{4}\; \Big(2\; (\alpha(\beta))^{2} - 2\; \alpha(\beta) + 4 + D
                 + 3 (2\; (\alpha(\beta))^{2} - 4\; \alpha(\beta) \nn \\
            & &  + 8 + D) \; \Upsilon_{BG}(\beta) + 4\; ((\omega(\beta))^{2} + (\alpha(\beta) - 3) \; \omega(\beta)
                 - (\alpha(\beta) - 2))\; \ln \Upsilon_{BG}(\beta) \nn \\
            & &  + \omega(\beta)\; (1 + \omega(\beta))\; ((\alpha(\beta))^{2} - \alpha(\beta) + 2)\Big)
                   V_{BG}(\beta) + \cdots,
\label{avg_vol_GE}
\eea
where $V_{BG}(\beta)$ is the average volume in the Boltzmann-Gibbs formalism. Similarly the average number of particles
can be arrived through the use of the relation $N_{q}^{(3)} = \frac{\partial {\mathcal{E}}_{q}^{(3)}}{\partial \mu}$.
The final expression of the average number as a perturbative series upto first order in $(1-q)$ is
\bea
N_{q}^{(3)} &=&   N_{BG}(\beta) + (1-q) \; \frac{1}{8}\; \bigg(\big(4 \;(\beta \mu)^{2} + 8\,  \beta P \, \alpha(\beta)
                - D(2+D) + 8\, (\alpha(\beta)
\label{avg_no_GE}  \\
            & & + 1) \ln \Upsilon_{BG}(\beta)\big) \; N_{BG}(\beta) - \big(12 \,(\beta \mu)^{2} - 4D\, \beta \mu - 8\, \beta P
                  (\alpha(\beta) + 1) - D(2+D) \nn \\
            & & + 8\, (\alpha(\beta) - 1) \ln \Upsilon_{BG}(\beta)\big) \; (N_{BG}(\beta))^{2}
                + 4\, (2\, \beta\mu - \beta P - 1) \; (N_{BG}(\beta))^{3}\bigg) + \cdots.   \nn
\eea
Using the perturbative expansions of the heat function (\ref{Rhf_GE_Exp}), the average volume (\ref{avg_vol_GE})
and the average number of particles (\ref{avg_no_GE}), the internal energy can be computed via the
legendre transformation ${\mathsf{R}}_{q}^{(3)} = U_{q}^{(3)} + P V_{q}^{(3)} - \mu N_{q}^{(3)}$. The computed
perturbative series of the internal energy upto $(1-q)$ order reads:
\bea
U_{q}^{(3)} &=&   U_{BG}(\beta) - (1-q)\; \frac{1}{8} \; \Upsilon_{BG}(\beta) \big((8\; \ln \Upsilon_{BG}(\beta)
                + \alpha(\beta) - 1) \; \omega(\beta) + 8\; (\alpha(\beta) \nn \\
            & & - 2) \ln \Upsilon_{BG}(\beta) + 4\, (\beta \mu - 2) \alpha(\beta) + 2D \, \beta \mu - (D+1)^{2}\big)\, U_{BG}(\beta)
                + \cdots.
\label{avg_Ie_GE}
\eea
The quantities evaluated above can also be arrived at using the basic definition of the
internal energy, the average volume and the average number of particles.  For this purpose we employ
the basic definition of these quantities to find the corresponding implicit equations
which are then solved recursively to obtain the final explicit expressions. The results
obtained through this method perfectly agrees with the results displayed above. All the
quantities recover the corresponding Boltzmann-Gibbs results in the $q \rightarrow 1$
limit.
\subsection{Extreme relativistic ideal gas}
The extreme relativistic ideal gas has been studied using the perturbative formalism. The actual process of
calculation is same as in the nonrelativistic case and so we present the final results directly.  The
perturbative expansion corresponding to the the average number of particles is
\bea
\mathtt{N}_{q}^{(3)} &=& \mathtt{N}_{BG} - \frac{(1-q)}{2} \ \varUpsilon_{BG}(\beta) \;
                         \Big(D + D^{2} - 2D \, \beta \mu +(\beta \mu)^{2} \nn \\
                     & & - 2\, (1+D-\beta \mu)\; \ln \varUpsilon_{BG}(\beta)\Big) \; \mathtt{N}_{BG}(\beta)
                         + \cdots.
\label{avgN_GE_ee}
\eea
The factors $\varUpsilon_{BG}(\beta)$ and $\mathtt{N}_{BG}(\beta)$, are the partition function and average number in the Boltzmann-Gibbs statistics respectively and their expressions are given below
\beq
\varUpsilon_{BG}(\beta) = \frac{1}{1-\xi(\beta)}, \qquad
\mathtt{N}_{BG} = \frac{\xi(\beta)}{1-\xi(\beta)}, \qquad
\xi(\beta) = \exp(\beta \mu)\;  \Delta \;  \frac{1}{P} \; \left(\frac{1}{\beta}\right)^{D+1}
\eeq
is the average number in the Boltzmann-Gibbs statistics and $\xi(\beta)$ is assumed to be less than 1.  The average volume
in nonextensive statistical mechanics is
\bea
\mathtt{V}_{q}^{(3)} &=& \mathtt{V}_{BG} - \frac{(1-q)}{2}\; \varUpsilon_{BG}(\beta)
                         \Big(\big(2 + 3D + D^{2} - 2(1+D) \beta \mu - (\beta \mu)^{2}  \\
                     & & - 2  (2 + D - \beta \mu) \ln \varUpsilon_{BG}(\beta)\big) \mathtt{V}_{BG}(\beta)
                         - 2  (1 + D - \beta \mu - \ln \varUpsilon_{BG}) (\mathtt{V}_{BG}(\beta))^{2}\Big) + \cdots. \nn
\label{avgVol_GE_ee}
\eea
where the Boltzmann-Gibbs value corresponding to it is $\mathtt{V}_{BG} = \frac{1}{\beta\, P} \; \mathtt{N}_{BG}(\beta)$.
A similar perturbative series corresponding to the internal energy is
\bea
\mathtt{U}_{q}^{(3)} &=& \mathtt{U}_{BG} - \frac{(1-q)}{2}\; \varUpsilon_{BG}(\beta)
                         \Big(\big(2 + 3D + D^{2} - (1+D) \beta \mu - (\beta \mu)^{2} \\
                     & & - 2  (2 + D - \beta \mu) \ln \varUpsilon_{BG}(\beta)\big)\mathtt{U}_{BG}(\beta)
                         - 2  (1 + D - \beta \mu - \ln \varUpsilon_{BG}) (\mathtt{U}_{BG}(\beta))^{2}\Big) + \cdots. \nn
\label{avgIE_GE_ee}
\eea
The Boltzmann-Gibbs internal energy used in (\ref{avgIE_GE_ee}) is $\mathtt{U}_{BG}(\beta)
= \frac{D}{\beta} \; \mathtt{N}_{BG}(\beta)$.  From the results displayed above we can observe that in the
$q \rightarrow 1$ limit we recover the standard Boltzmann-Gibbs values.

\par

The Tonks gas described by the Hamiltonian (\ref{ham_tg}) can also be studied through the generalized ensemble. The
results corresponding to such a model of interacting gas in the free length generalized ensemble can be obtained by
substituting $D=1$ in the results obtained for the nonrelativistic classical ideal gas.
%
%
%
%
\setcounter{equation}{0}
\section{Adiabatic ensemble}
\label{Adiabatic}
The adiabatic class of ensembles is used to describe a system of particles for
which the thermal equilibration is with respect to the heat function.
In the adiabatic class the individual members of the ensemble
have the same value of the heat function though they can be at different temperatures.
There are four different adiabatic ensembles namely microcanonical ($N$, $V$, $U$),
isoenthalpic-isobaric ensemble ($N$, $P$, $H$), the ensemble with number fluctuations
($\mu$, $V$, $\mathsf{L}$), and the ensemble with both the number and volume fluctuations
($\mu$, $P$, $\mathsf{R}$).  In the current section we present a unified framework to
describe the adiabatic class of ensembles.

\par

At any given instant in time the microstate of a system consisting of $N$ particles is represented
by a point in the $6N$ dimensional phase space, which is comprised of $3N$ position coordinates
and the $3N$ momentum coordinates. Since the position and momentum coordinates evolve with time,
the representative point moves in the phase space. The motion of the representative point traces a
trajectory of constant heat function ${\mathfrak{H}}$ in the phase space. The points which lie on the
surface of the constant heat function curve are the various microstates corresponding to the
macrostate of constant heat function.  As the number of microstates is very high and lie very close
to each other, the surface area of the constant heat function curve can be considered as a measure
of the total number of microstates. The surface area of the constant heat function ${\mathfrak{H}}$
curve in an adiabatic ensemble is
\bea
\Omega (\mathfrak{X}_{1},\mathfrak{X}_{2},\mathfrak{H})
              = \sum_{X_{\{\ell\}}} \, \frac{1}{N! \, h^{DN}}
                \int_{r_{i}} \int_{p_{i}}
                \delta \Big(\mathcal{H}+\sum_{\{\ell\}}
                x_{\ell}X_{\ell} - {\mathfrak{H}}\Big)
                \prod_{i=1}^{N} {\rm d}^{D} r_{i} \, {\rm d}^{D} p_{i}.
\label{sds_en}
\eea
The computation of area of the constant heat function is difficult, and so we usually calculate
the volume enclosed by such curve.  The phase space volume enclosed by the constant heat function
${\mathfrak{H}}$ curve is
\beq
\varSigma(\mathfrak{X}_{1},\mathfrak{X}_{2},\mathfrak{H}) =
                   \sum_{X_{\{\ell\}}} \, \frac{1}{ N! \, h^{DN}}
                   \int_{r_{i} } \int_{p_{i}}
                   \Theta \Big(\mathcal{H} + \sum_{\{\ell\}}
                   x_{\ell}X_{\ell} - {\mathfrak{H}} \Big)
                   \prod_{i=1}^{N} {\rm d}^{D} r_{i}\,  {\rm d}^{D} p_{i}.
\label{vds_en}
\eeq
The surface area of the heat function curve (\ref{sds_en}) and the volume (\ref{vds_en}) enclosed by it are related via the
relation
\beq
\varSigma(\mathfrak{X}_{1},\mathfrak{X}_{2},\mathfrak{H}) = \frac{\partial}{\partial \mathfrak{H}}\,
                                                            \Omega(\mathfrak{X}_{1},\mathfrak{X}_{2},\mathfrak{H}).
\label{sds_vds_rel}
\eeq
Based on the kind of adiabatic confinement there are four different definitions of the entropy.  A unified
definition of the entropy for an adiabatic nonextensive system is
\beq
S_{q}(\mathfrak{X}_{1},\mathfrak{X}_{2},\mathfrak{H}) = k \; \ln_{q} \varSigma(\mathfrak{X}_{1},\mathfrak{X}_{2},\mathfrak{H}),
\label{ent_iee}
\eeq
where the $\varSigma(\mathfrak{X}_{1},\mathfrak{X}_{2},\mathfrak{H})$ is the volume enclosed by the curve of constant
heat function and a measure of the number of microstates. The temperature of a general adiabatic ensemble is
defined via the relation
\beq
T = \left(\frac{\partial S_{q}}{\partial {\mathfrak{H}}}\right)^{-1}
  = \frac{(\varSigma(\mathfrak{X}_{1},\mathfrak{X}_{2},\mathfrak{H}))^{{}^{q}}}{k  \;\Omega(\mathfrak{X}_{1},\mathfrak{X}_{2},\mathfrak{H})}.
\label{temp_def}
\eeq
In the usual procedure of calculation the heat function can be computed through
the use of (\ref{temp_def}) and the relations (\ref{sds_en}) and (\ref{vds_en}).
Using the heat function the specific heat can calculated through the expression
\beq
C_{q}\big|_{x_{\{\ell\}}} = \frac{\partial {\mathfrak{H}}_{q}}
                 {\partial T} \bigg|_{x_{\{\ell\}}}.
\label{psh_def}
\eeq
The extensive thermodynamic variable whose intensive counterparts are held fixed can
be obtained from the relation
\beq
X_{\ell:q}(\mathfrak{X}_{1},\mathfrak{X}_{2},\mathfrak{H}) = - \frac{1}{\beta} \;
                                                               \frac{\partial}{\partial x_{\ell}}\,
                                                               S_{q}(\mathfrak{X}_{1},\mathfrak{X}_{2},\mathfrak{H}).
\label{vol_N_def}
\eeq
The volume and the average number of particles can be calculated in the ($N$, $P$, $H$) and
($\mu$, $V$, $\mathsf{L}$) respectively through the use of (\ref{vol_N_def}).  In the case of
($\mu$, $P$, $\mathsf{R}$) ensemble, both the volume and the average number of particles can be
calculated using (\ref{vol_N_def})
The expectation value of an observable $O$  for any adiabatic ensemble can be
defined in a unified way through the relation
\beq
\langle O \rangle = \frac{1}{\Omega(\mathfrak{X}_{1},\mathfrak{X}_{2},\mathfrak{H})} \,
                    \sum_{X_{\{\ell\}}}
                    \frac{1}{N! \, h^{DN}} \,
                    \int_{r_{i} } \int_{p_{i}} O \;\;
                    \delta \Big(\mathcal{H} + \sum_{\{\ell\}}
                    x_{\ell} X_{\ell} - {\mathfrak{H}}\Big)
                    \prod_{i=1}^{N} {\rm d}^{D} r_{i} \, {\rm d}^{D} p_{i}.
\label{oex_sds}
\eeq
The internal energy is the natural heat function of the microcanonical ensemble and so
it can be directly evaluated, whereas in the other three ensembles, equation (\ref{oex_sds})
is used to calculate the internal energy.  From the internal
energy the specific heat at constant volume can be evaluated using
\beq
C_{q}\big|_{V} = \frac{\partial U_{q}}
                 {\partial T} \bigg|_{V}.
\label{sph_cv}
\eeq
A nontrivial way of checking the internal energy can be accomplished by using the Legendre transform
(\ref{hf_def}) to find the internal energy.

\par

A $q$ generalization of the equipartition theorem is developed in the unified framework. Let $y_{i}$ be a phase space
variable which can denote either the position $r_{i}$ or the momentum $p_{i}$ $(i = 1,...,3N)$ coordinate.
Calculating the expectation value of $y_{i} \; \frac{\partial \mathcal{H}}{\partial y_{j}}$ we arrive at
\beq
\bigg\langle y_{i}\; \frac{\partial {\mathcal{H}}}
{\partial y_{j}} \bigg\rangle = k T \;
                                 \left(\varSigma(\mathfrak{X}_{1},\mathfrak{X}_{2},\mathfrak{H})\right)^{q-1} \; \delta_{ij}.
\label{ept_uf}
\eeq
Equation (\ref{ept_uf}) obtained above is the generalized form of the $q$-nonextensive equipartition theorem.
It is interesting to note that the nonextensive generalization of the equipartition theorem depends on the factor $\varSigma(\mathfrak{X}_{1},\mathfrak{X}_{2},\mathfrak{H})$ which is a measure of the number of microstates
corresponding to a given macrostate. However we notice that in the extensive $q \rightarrow 1$ limit the
dependence on the number of microstates vanishes.  The form of (\ref{ept_uf}) suggests that the
the expectation value described has a nonzero value only when $i=j$.  When the phase space variable $y_{i}$ is set to be
the coordinate $r_{i}$ then we get a specific form of the equipartition theorem
\beq
\bigg\langle r_{i}\; \frac{\partial {\mathcal{H}}}
{\partial r_{i}} \bigg\rangle = - \langle r_{i} \;  \dot{p}_{i}\rangle =  \langle r_{i} \; F_{i} \rangle = k T \;
                                 \left(\varSigma(\mathfrak{X}_{1},\mathfrak{X}_{2},\mathfrak{H})\right)^{q-1},
\label{ept_uf_csp}
\eeq
where we have used  Hamilton's equation of motion $\frac{\partial \mathcal{H}}{\partial r_{i}} = - \dot{p}_{i}$.
Following an identical approach for the situation where the generalized variable is set to be the momentum variable
$y_{i} = p_{i}$ we arrive at
\beq
\bigg\langle p_{i}\; \frac{\partial {\mathcal{H}}}
{\partial p_{i}} \bigg\rangle = \langle p_{i}\; \dot{q}_{i} \rangle = k T \;
                                 \left(\varSigma(\mathfrak{X}_{1},\mathfrak{X}_{2},\mathfrak{H})\right)^{q-1}.
\label{ept_uf_msp}
\eeq
We observe that (\ref{ept_uf_msp}) is nothing but twice the expectation value of kinetic energy.
From (\ref{ept_uf_csp}) and (\ref{ept_uf_msp}) it can be proved that systems whose Hamiltonians can be cast in
the form $\mathcal{H} = \sum_{i} A_{i}\, P_{i}^{2} + \sum_{i} B_{i}\, Q_{i}^{2}$  through a canonical transformation
the expectation value of the Hamiltonian is
\beq
\mathcal{H} = \frac{D N}{2} \;  k T \; \left(\varSigma(\mathfrak{X}_{1},\mathfrak{X}_{2},\mathfrak{H})\right)^{q-1}.
\label{ept_qh}
\eeq
The virial theorem in the nonextensive framework obtained from (\ref{ept_uf_csp}) reads:
\beq
\bigg\langle \sum_{i}^{3N} q_{i} \, \dot{p}_{i} \bigg\rangle = - 3N\; k T \;
                                                                 \left(\varSigma(\mathfrak{X}_{1},\mathfrak{X}_{2},
                                                                 \mathfrak{H})\right)^{q-1}.
\label{vt_next}
\eeq
The unified framework derived above is made use of in the following sections to demonstrate the
isoenthalpic-isobaric ensemble ($N$,$P$,$H$), the ensemble with number fluctuations ($\mu$,$V$,$\mathsf{L}$)
and the ensemble with both number and volume fluctuations ($\mu$,$P$,$\mathsf{R}$).  Though all the
relations in the microcanonical ensemble ($N$,$V$,$U$) can be read from the equations in the unified
framework above, we do not discuss the ($N$,$V$,$U$) ensemble, since a detailed study has already been
carried out in [\cite{CVG}].
%
%
%
%
\setcounter{equation}{0}
\section{Isoenthalpic-Isobaric ensemble}
\label{NPH}
The study of adiabatically confined systems with constant enthalpy is
done using the Isoenthalpic-Isobaric ensemble. In the following discussion we
study the classical nonrelativistic and extreme relativistic ideal gas.
\subsection{Nonrelativistic ideal gas in $(N,P,H)$ ensemble}
The nonrelativistic ideal gas is studied in the isoenthalpic-isobaric ensemble. The
volume enclosed by the surface of constant enthalpy is calculated in the following
manner.  First the Hamiltonian corresponding to the nonrelativisitic gas (\ref{Ham_nr})
is substituted in equation (\ref{vds_en}) and the resultant expression reads:
\beq
\varSigma(N,P,H) = \sum_{V} \frac{1}{N! \, h^{DN}} \; \int_{r_{i}} \int_{p_{i}}
                   \Theta\left(\sum_{i} \frac{p_{i}^{2}}{2m} + P V - H\right)\;
                   \prod_{i=1}^{N} {\rm d}^{D} r_{i}\, {\rm d}^{D} p_{i}.
\label{VS_nrg}
\eeq
The momentum integration in (\ref{VS_nrg}) is geometrically the volume of $DN$
dimensional sphere of radius $\sqrt{2m\, (H - PV)}$ and is equal to
\beq
{}_{nr} \mathcal{I}_{p} = \int_{\sum_{i}^{DN} \frac{p_{i}^{2}}{2m} \leq H - P V}
                          {\rm d}^{DN} p_{i} = \frac{(2\pi m)^{\frac{DN}{2}}}
                          {\Gamma\left(\frac{D N}{2}+1\right)} \,(H-P V)^{\frac{DN}{2}}.
\label{p_int}
\eeq
Substituting the result of the momentum integration (\ref{p_int}) in (\ref{VS_nrg}) the
expression for the volume is
\beq
\varSigma(N,P,H) = \frac{\mathcal{M}^{N}}{N!} \; \frac{1}{\Gamma\left(\frac{D N}{2}+1\right)} \;
                   \sum_{V} (H-P V)^{\frac{DN}{2}} \int {\rm d}^{DN} r_{i}.
\label{r_int}
\eeq
In the next step the integral over the position coordinates is carried out in (\ref{r_int})
and this yields
\beq
\varSigma(N,P,H) = \frac{\mathcal{M}^{N}}{N!} \; \frac{1}{\Gamma\left(\frac{D N}{2}+1\right)} \;
                   \sum_{V} V^{N} (H-P V)^{\frac{DN}{2}}.
\label{V_sum}
\eeq
Finally the summation over the volume eigenstates is considered.  Since the volume states are
very closely spaced the summation is approximated by an integral.  But as we have discussed
earlier an integration leads to overcounting of volume states.  To overcome this we employ
the shell particle method of counting of volume states proposed in [\cite{DC1},\cite{DC2}],
wherein only the distinct minimum volume states needed to confine the particle is
taken into account.  The volume enclosed by the isoenthalpic curve in phase space is
\beq
\varSigma(N,P,H) = \mathcal{M}^{N} \;
               \left(\frac{1}{P}\right)^{N} \;
               \frac{H^{{\mathfrak{D}}_{1}}}
               { \Gamma( {\mathfrak{D}}_{1} + 1 ) }
\label{vds_nr_fe}
\eeq
The surface area of the constant enthalpy curve in the phase space is
\beq
\Omega(N,P,H) = \mathcal{M}^{N} \;
            \left(\frac{1}{P}\right)^{N} \;
            \frac{H^{{\mathfrak{D}}_{1}-1}}
            { \Gamma( {\mathfrak{D}}_{1}) }
\label{sds_nr_fe}
\eeq
From the volume of the isoenthalpic curve in the phase space the nonextensive entropy of the
classical ideal gas in this ensemble is
\beq
S_{q}(N,P,H) = k \left(\frac{\left(\mathcal{M}^{N} \;
               \left(\frac{1}{P}\right)^{N} \;
               \frac{H^{{\mathfrak{D}}_{1}}}
               {\Gamma( {\mathfrak{D}}_{1} + 1)}\right)^{1-q}-1} {1-q}\right).
\label{ent_erg_fe}
\eeq
Using the definition of temperature (\ref{temp_def}), the heat function corresponding to the
ensemble, the enthalpy is found
\beq
H_{q} = {\mathcal{M}}^{(1-q)\, N {\mathfrak{L}}_{1}} \;
    \left(\frac{({\mathfrak{D}}_{1})^{q}}
    {\left(\Gamma({\mathfrak{D}}_{1})\right)^{1-q}}\right)
    ^{{\mathfrak{L}}_{1}} \;
    \left(\frac{1}{P}\right)^{(1-q)N {\mathfrak{L}}_{1}} \;
    \left(\frac{1}{\beta}\right)^{{\mathfrak{L}}_{1}}
\label{en_fe}
\eeq
The specific heat at constant pressure evaluated from (\ref{en_fe}) through the
use of (\ref{psh_def}) is
\beq
C_{q} \big|_{P} = \left({\mathfrak{D}}_{1}\right)^{q {\mathfrak{L}}_{1}}
                  \; k \; {\mathfrak{L}}_{1} \;
                  \left(\frac{\mathcal{M}^{N}}{\Gamma({\mathfrak{D}}_{1})} \;
                  \left(\frac{1}{P}\right)^{N} \;
                  \left(\frac{1}{\beta}\right)^{\mathfrak{D}_{1}}\right)^
                  {(1-q) {\mathfrak{L}}_{1}}.
\label{psh_fe}
\eeq
The average volume calculated using (\ref{vol_N_def}) can be rewritten to obtain the
equation of state given below
\beq
P V_{q} = \frac{N}{\beta} \left(\mathcal{M}^{N} \;
     \frac{({\mathfrak{D}}_{1})^{{\mathfrak{D}}_{1}}}
     {\Gamma({\mathfrak{D}_{1}})} \;
     \left(\frac{1}{P}\right)^{N} \;
     \left(\frac{1}{\beta}\right)^{{\mathfrak{D}}_{1}}\right)^
     {(1-q) {\mathfrak{L}}_{1}}.
\label{eqs_fe}
\eeq
The internal energy evaluated through the definition of the
expectation value of the observable (\ref{oex_sds}) is
\beq
U_{q}(\beta)  = \frac{DN}{2 \beta} \; \left(\frac{\mathcal{M}^{N}}
                {\Gamma({\mathfrak{D}}_{1}+1)} \;
                \left(\frac{1}{P}\right)^{N} \;
                \left(\frac{1}{\beta}\right)^{\mathfrak{D}_{1}}\right)^
                {(1-q) {\mathfrak{L}}_{1}}.
\label{ie_IeIb}
\eeq
The calculated value of the internal energy has also been checked through a
different procedure using the Legendre transformation $H_{q} = U_{q} + P V_{q}$.
From (\ref{ie_IeIb}) the specific heat at constant volume found using (\ref{sph_cv}) is
\beq
C_{q} \big|_{V} = \frac{DN}{2} \; k \; {\mathfrak{L}}_{1} \;
                  \left(\frac{\mathcal{M}^{N}}{\Gamma({\mathfrak{D}}_{1}+1)} \;
                  \left(\frac{1}{P}\right)^{N} \;
                  \left(\frac{1}{\beta}\right)^{\mathfrak{D}_{1}}\right)^
                  {(1-q) {\mathfrak{L}}_{1}}.
\label{vsh_fe}
\eeq
Using the results corresponding to the specific heat at constant pressure
(\ref{psh_fe}) and the specific heat at constant volume (\ref{vsh_fe})
the $q$-generalization of the Mayer's relation is obtained
\beq
C_{q} \big|_{P} - C_{q} \big|_{V} = N  k \; {\mathfrak{L}}_{1} \;
                                    \left(\frac{\mathcal{M}^{N}}
                                    {\Gamma({\mathfrak{D}}_{1}+1)} \;
                                    \left(\frac{1}{P}\right)^{N} \;
                                    \left(\frac{1}{\beta}\right)
                                    ^{\mathfrak{D}_{1}}\right)^
                                    {(1-q) {\mathfrak{L}}_{1}}.
\label{My_rel_IB}
\eeq
The ratio between the specific heat at constant pressure (\ref{psh_fe})
and the specific heat at constant volume (\ref{vsh_fe})
\beq
\gamma = \frac{C_{q} \big|_{P}}{C_{q} \big|_{V}} = 1 + \frac{2}{D}
\label{sph_ratio}
\eeq
is interestingly independent of $q$ and $N$. We have observed the same result in the isothermal-isobaric
ensemble (\ref{poly_ind}).  In the extensive $q \rightarrow 1$ limit the standard
Boltzmann-Gibbs results are recovered.
\subsection{Extreme relativistic ideal gas in $(N,P,H)$ ensemble}
The extreme relativistic classical ideal gas described by the Hamiltonian (\ref{Ham_er})
is studied in the $(N,P,H)$ ensemble. The volume enclosed by the surface of constant enthalpy is
given by
\beq
\varSigma(N,P,H) = \sum_{V} \frac{1}{N! \, h^{DN}} \; \int_{r_{i}} \int_{p_{i}}
                   \Theta\Big(c \, \sum_{i} p_{i} + PV - H\Big)\;
                   \prod_{i=1}^{N} {\rm d}^{D} r_{i}\, {\rm d}^{D} p_{i}.
\label{VS_erg}
\eeq
In (\ref{VS_erg}) the result of the momentum integration is
\beq
{}_{er} \mathcal{I}_{p} = \int_{c \sum_{i}^{DN} p_{i} \leq H - PV}   {\rm d}^{DN} p_{i}
                        =  \frac{1}{\Gamma\left(DN + 1\right)}\;
                          \left(\frac{2 \pi^{\frac{D}{2}} \; \Gamma(D)}{\Gamma\left(\frac{D}{2}\right)}\right)^{N} \,(H-PV)^{DN}.
\label{p_int_er}
\eeq
The expression (\ref{p_int_er}) is substituted in (\ref{VS_erg}) and the resulting expression is integrated over the position
coordinates.  Replacing the summation over the volume eigenstates by an integration to consider the continuum nature of
volume and adopting the shell particle method of counting of volume states the final expression for $\varSigma(N,P,H)$ is
\beq
\varSigma(N,P,H) =  \Delta^{N} \; \left(\frac{1}{P}\right)^{N} \;
               \frac{H^{{\mathfrak{D}}_{2}}}
               { \Gamma({\mathfrak{D}}_{2} + 1) },
\label{vds_er_fe}
\eeq
which enables us to calculate the surface of the isoenthalpic curve via the relation (\ref{sds_vds_rel})
\beq
\Omega(N,P,H) = \Delta^{N} \; \left(\frac{1}{P}\right)^{N} \;
            \frac{H^{{\mathfrak{D}}_{2}-1}}
            {\Gamma({\mathfrak{D}}_{2})}.
\label{sds_er_fe}
\eeq
The nonextensive entropy (\ref{ent_iee}) of the extreme relativistic ideal gas obtained from
(\ref{vds_er_fe}) is
\beq
S_{q}(N,P,H) = k \left(\frac{\left(\Delta^{N} \; \left(\frac{1}{P}\right)^{N} \;
               \frac{H^{{\mathfrak{D}}_{2}}}
               {\Gamma({\mathfrak{D}}_{2} + 1)}\right)^{1-q}-1} {1-q}\right).
\label{ent_fe}
\eeq
Using the definition of temperature in the isoenthalpic-isobaric ensemble (\ref{temp_def}) in conjunction with the
expression for the entropy (\ref{ent_fe}) we find the enthalpy
\beq
H_{q} = \Delta^{(1-q) N {\mathfrak{L}}_{2}} \;
        \bigg( \frac{ ( {\mathfrak{D}}_{2})^{q} }
        {\left( \Gamma( {\mathfrak{D}}_{2} ) \right)^{1-q} } \bigg)
        ^{{\mathfrak{L}}_{2}} \;
        \bigg( \frac{1}{P} \bigg)^{(1-q)N {\mathfrak{L}}_{2}} \;
        \bigg( \frac{1}{\beta} \bigg)^{{\mathfrak{L}}_{2}}.
\label{en_fe_er}
\eeq
The specific heat at constant pressure obtained from the enthalpy via
(\ref{psh_def}) is
\beq
C_{q} \big|_{P} = \left({\mathfrak{D}}_{2}\right)^{q {\mathfrak{L}}_{2}}
                  \; k \; {\mathfrak{L}}_{2} \;
                  \left(\frac{\Delta^{N}}{\Gamma({\mathfrak{D}}_{2})} \;
                  \left(\frac{1}{P}\right)^{N} \;
                  \left(\frac{1}{\beta}\right)^{\mathfrak{D}_{2}}\right)^
                  {(1-q) {\mathfrak{L}}_{2}}.
\label{psh_fe_er}
\eeq
The computed expression for the average volume using (\ref{vol_N_def}) is rewritten
to give the equation of state
\beq
P V_{q} = \frac{N}{\beta} \left(\Delta^{N} \;
          \frac{({\mathfrak{D}}_{2})^{{\mathfrak{D}}_{2}}}
          {\Gamma({\mathfrak{D}_{2}})} \;
          \left(\frac{1}{P}\right)^{N} \;
          \left(\frac{1}{\beta}\right)^{{\mathfrak{D}}_{2}}\right)^
          {(1-q) {\mathfrak{L}}_{2}}.
\label{eqs_fe_er}
\eeq
Making use of the formula for the expectation value (\ref{oex_sds}), the calculated form of the
internal energy is
\beq
U_{q}(\beta)  = \frac{D N}{\beta} \; \left(\frac{\Delta^{N}}
                {\Gamma({\mathfrak{D}}_{2}+1)} \;
                \left(\frac{1}{P}\right)^{N} \;
                \left(\frac{1}{\beta}\right)^{\mathfrak{D}_{2}}\right)^
                {(1-q) {\mathfrak{L}}_{2}},
\label{ie_IeIb_er}
\eeq
and this perfectly agrees with the result found using the Legendre transformation $H=U+PV$.
From the internal energy the specific heat at constant volume can be found:
\beq
C_{q} \big|_{V} = D N \; k \; {\mathfrak{L}}_{2} \;
                  \left(\frac{\Delta^{N}}{\Gamma({\mathfrak{D}}_{2}+1)} \;
                  \left(\frac{1}{P}\right)^{N} \;
                  \left(\frac{1}{\beta}\right)^{\mathfrak{D}_{2}}\right)^
                  {(1-q) {\mathfrak{L}}_{2}}.
\label{vsh_fe_er}
\eeq
The ratio between the heat capacities at constant pressure and constant volume
\beq
\gamma = \frac{C_{q} \big|_{P}}{C_{q} \big|_{V}} = 1 + \frac{1}{D},
\label{sph_ratio_erg}
\eeq
is found to be independent of both the nonextensive parameter and the number of particles $N$, similar to
the result observed in (\ref{sh_rat_erg}).
The enthalpy, internal energy and the specific heats go to their respective Boltzmann-Gibbs value
in the $q \rightarrow 1$ limit.

\par

The thermodynamic quantities corresponding to the Tonks gas (\ref{ham_tg}) can also be
computed through a procedure identical to the one employed in the case of the
nonrelativistic and extreme relativistic classical ideal gas.  The results can be easily read
from the expressions corresponding to the nonrelativistic gas through the use of the free length
$L_{f:q}$ in the place of the volume $V$ and setting the dimension $D=1$. The equation of
state of the Tonks gas thus obtained is
\beq
P L_{f:q} =  \frac{N}{\beta} \; {\mathcal{M}}_{1}^{(1-q) N
            {\mathfrak{L}}_{1:1}} \; \left(\frac{\frac{3N}{2}}
            {\Gamma\left(\frac{3N}{2}\right)}\right)
            ^{(1-q)\frac{3N}{2}{\mathfrak{L}}_{1:1}}
            \left(\frac{1}{P}\right)^{(1-q)N
            {\mathfrak{L}}_{1:1}} \; \left(\frac{1}{\beta}
            \right)^{(1-q)\frac{3N}{2}{\mathfrak{L}}_{1:1}}.
\label{eos_tg}
\eeq
It can be seen that equation (\ref{eos_tg}) recovers its standard Boltzmann-Gibbs result
$P L_{f:q} = N\, kT$ in the $q \rightarrow 1$ limit.
%
%
%
%
\setcounter{equation}{0}
\section{Adiabatic ensemble with number fluctuations}
\label{LVmu}
The adiabatic ensemble with number fluctuations was first discussed in detail in
[\cite{JR2}] in the context of Boltzmann-Gibbs ensemble. This ensemble is the adiabatic counterpart
of the grandcanonical ensemble and the Hill energy is the corresponding heat function.
Though the integration with respect to the position and the momentum coordinates can be carried easily,
it is not possible to carry the sum over all possible values of $N$. So in this section we give only a
formal expression corresponding to the phasespace volume, surface area, the Hill energy and the
equation of state.
\subsection{Nonrelativistic Ideal gas}
The volume of phasespace enclosed by the curve of constant Hill energy $\mathsf{L}$ after carrying out
the integrations over the position and the momentum coordinates is
\beq
\varSigma(\mu,V,{\mathsf{L}}) = \sum_{N=0}^{\infty} \;
                                       \frac{V^{N}}{N!\; {\Gamma\left(\frac{D N}{2}+1\right)}} \;\;
                                       {\mathcal{M}}^{N} \;\;
                                      ({\mathsf{L}}+\mu N)^{\frac{DN}{2}}.
\label{psv_Lmu_nr}
\eeq
The surface area corresponding to the phasespace curve of constant Hill energy can arrived at
through the use of (\ref{sds_vds_rel}). The equation of state describing the classical ideal
gas in this ensemble is
\beq
P V_{q} = \frac{k T}{\varSigma^{{}^{q}}} \;\;
     \sum_{N=0}^{\infty} N  \; \frac{V^{N}}{N! \; \Gamma\left(\frac{D N}{2}+1\right)} \;\;
     {\mathcal{M}}^{N} \;\; ({\mathsf{L}}+\mu N)^{\frac{DN}{2}}.
\label{Eos_Lmu_nr}
\eeq
The formal expression for the average number of particles arrived at through the use of
(\ref{vol_N_def}) is
\beq
N_{q} = \frac{kT}{\varSigma^{{}^{q}}} \;\;
    \sum_{N=0}^{\infty} N \; \frac{V^{N}}{N! \; \Gamma\left(\frac{DN}{2}\right)} \;\;
    {\mathcal{M}}^{N} \;\;({\mathsf{L}}+\mu N)^{\frac{DN}{2}-1}.
\label{Navg_Lmu_nr}
\eeq
\subsection{Extreme relativistic Ideal gas}
For an extreme relativistic classical ideal gas described by the hamiltonian
(\ref{Ham_er}), the phase space volume calculated is
\beq
\varSigma(\mu,V,{\mathsf{L}}) = \sum_{N=0}^{\infty} \;
                                 \frac{V^{N}}{N!\; {\Gamma\left(DN+1\right)}} \;\;
                                 \Delta^{N} \;\;
                                 ({\mathsf{L}}+\mu N)^{DN}.
\label{psv_Lmu_er}
\eeq
The equation of state and the average number of particles of the extreme relativistic
gas corresponding to this adiabatic ensemble are
\bea
P V_{q} &=& \frac{k T}{\varSigma^{{}^{q}}} \;\;
     \sum_{N=0}^{\infty} N  \; \frac{V^{N}}{N! \; \Gamma\left(DN\right)} \;\;
     \Delta^{N} \;\; ({\mathsf{L}}+\mu N)^{DN},
\label{Eos_Lmu_er} \\
N_{q} &=& \frac{k T}{\varSigma^{{}^{q}}} \;\;
    \sum_{N=0}^{\infty} N \; \frac{V^{N}}{N! \; \Gamma\left(DN-1\right)} \;\;
    \Delta^{N} \;\;({\mathsf{L}}+\mu N)^{DN-1}.
\label{Navg_Lmu_er}
\eea
In the $q \rightarrow 1$ the above quantities reduce to the standard Boltzmann-Gibbs
equations. Substituting $D=1$ and replacing the volume by the free length in the thermodynamic
expression for the nonrelativistic ideal gas, we can arrive at relations corresponding to the Tonks gas.
%
%
%
%
\setcounter{equation}{0}
\section{Adiabatic ensemble with volume and \\
          number fluctuations}
\label{RPmu}
An adiabatically confined system which can exchange both volume and particles
with the bath is described by this ensemble.  In the classical Boltzmann-Gibbs
case such an ensemble was first described and discussed in detail in [\cite{JR4}].
The heat function of the system is the $\mathsf{R}$ heat function and this is the
adiabatic counterpart of the generalized ensemble. Similar to the previous ensemble, the
$N$ summation is difficult to carry and so we present the results as a formal sum.
\subsection{Nonrelativistic Ideal gas}
In the ensemble with both volume and number fluctuations, the phasespace volume
of the classical nonrelativistic ideal gas enclosed by a curve of constant $\mathsf{R}$
is
\beq
\varSigma(\mu,P,\mathsf{R}) = \sum_{N=0}^{\infty} \;
                                 \frac{\mathcal{M}^{N}}{{\Gamma\left({\mathfrak{D}}_{1}+1\right)}} \;\;
                                 \left(\frac{1}{P}\right)^{N}
                                 ({\mathsf{R}}+\mu N)^{{\mathfrak{D}}_{1}}.
\label{psv_Rmu_nr}
\eeq
Using the expression for the phase space volume (\ref{psv_Rmu_nr}), the equation of state
and the average number of particles is obtained
\beq
P V_{q} = \frac{k T}{\varSigma^{{}^{q}}} \;\;
     \sum_{N=0}^{\infty} N  \; \frac{{\mathcal{M}}^{N}}{\Gamma\left({\mathfrak{D}}_{1}+1\right)} \;\;
     \;\; ({\mathsf{R}}+\mu N)^{{\mathfrak{D}}_{1}},
\label{Eos_Rmu_nr}
\eeq
\beq
N_{q} = \frac{k T}{\varSigma^{{}^{q}}} \;\;
    \sum_{N=0}^{\infty} N \; \frac{{\mathcal{M}}^{N}}{\Gamma\left({\mathfrak{D}}_{1}\right)} \;\;
    ({\mathsf{R}}+\mu N)^{{\mathfrak{D}}_{1}-1}.
\label{Navg_Rmu_nr}
\eeq
\subsection{Extreme relativistic Ideal gas}
In the case of an extreme relativistic ideal gas, the phase space volume enclosed by a curve of constant
$\mathsf{R}$ is
\beq
\varSigma(\mu,P,\mathsf{R}) = \sum_{N=0}^{\infty} \;
                                 \frac{\Delta^{N}}{{\Gamma\left({\mathfrak{D}}_{2}+1\right)}} \;\;
                                 \left(\frac{1}{P}\right)^{N} \;\;
                                 ({\mathsf{R}}+\mu N)^{{\mathfrak{D}}_{2}},
\label{psv_Rmu_er}
\eeq
which leads to the equation of state and the average number of particles given below
\bea
P V_{q} &=& \frac{k T}{\varSigma^{{}^{q}}} \;\;
            \sum_{N=0}^{\infty} N  \; \frac{\Delta^{N}}{\Gamma\left({\mathfrak{D}}_{2}\right)} \;\;
            \left(\frac{1}{P}\right)^{N} \;\;
           ({\mathsf{R}}+\mu N)^{{\mathfrak{D}}_{2}},
\label{Eos_Rmu_er}  \\
N_{q} &=& \frac{k T}{\varSigma^{{}^{q}}} \;\;
      \sum_{N=0}^{\infty} N \; \frac{\Delta^{N}}{\Gamma\left({\mathfrak{D}}_{2}-1\right)} \;\;
      \left(\frac{1}{P}\right)^{N} \;\;
      ({\mathsf{R}}+\mu N)^{{\mathfrak{D}}_{2}-1}.
\label{Navg_Rmu_er}
\eea
The $q \rightarrow 1$ limit is respected in all the above mentioned cases.  The thermodynamic
expressions corresponding to the Tonks gas can be arrived directly from the relations given
for the nonrelativistic gas by substituting $D=1$ and replacing volume by the free length.

\par

The summation over the number of particles could not be evaluated in both the ensemble with the number
fluctuations and the ensemble with number and volume fluctuations.  But the thermodynamic
quantities in both these ensembles may be obtained through the use of a molecular dynamics
simulation.
%
%
%
%
\section{Remarks}
\label{remarks}
A comprehensive investigation on the different kinds of ensembles has been carried out
in the present work.  We notice that there are two different classes of ensembles namely
the isothermal class and the adiabatic class.  The thermal equilibration in the isothermal
class happens with respect to the temperature. Isothermal class comprises of the canonical ensemble,
the isothermal-isobaric ensemble, the grandcanonical ensemble and the generalized ensemble.
We notice that the thermodynamic relations corresponding to these four ensembles can be
presented through a unified formulation.  Such a formulation is evolved for both the second and the
third constraint formalisms though the latter is the currently accepted formulation of nonextensive
$q$-statistical mechanics. The development of unified formulation corresponding to the second
constraint has been done with a view to apply nonextensive $q$ statistics in computational simulations
like Montecarlo and molecular dynamics.  It is a well known fact that the probabilities in the third
constraint are implicit quantities. In calculational procedures involving simulations it is
easier to find the thermodynamic quantities in the second constraint and later transform them to the
third constraint. Towards this end a generalization of the temperature dependent interrelation between the
second and the third constraint introduced in Ref. [\cite{TMP}] has been accomplished.

\par

The unified formulation developed is applied to study the isothermal-isobaric, grandcanonical, and the generalized
ensembles through specific examples viz the classical nonrelativistic, and, extreme relativistic ideal gas models.
Since the canonical ensemble has already been studied in detail in Ref. [\cite{CUR},\cite{TMP}] it has not been considered
in our present work.  In the isothermal-isobaric ensemble, the implicit equations corresponding to the generalized
partition function and the enthalpy are solved to obtain their corresponding explicit expressions. From these
quantities the average volume, the internal energy, the specific heat at constant pressure, and, the specific heat
at constant volume are obtained. All the above mentioned thermodynamic quantities are obtained as exact results.
The specific heats were dependent on the nonextensivity parameter $q$ and the temperature in both the models.
Interestingly we notice that the ratio between the specific heat at constant pressure and the specific heat at
constant volume is independent of both $q$ and $N$.  A similar
exact computation of the gas models in the grandcanonical and the generalized ensembles were found to be difficult.
So, we employed a perturbative procedure developed in [\cite{CCN1},\cite{CCN2}] based on the disentangling of $q$-exponential [\cite{CQ}]. The perturbative series corresponding to the generalized partition function, the sum of $q$-weights and the heat function are obtained up to a predetermined order in the third constraint formalism. The determined implicit quantities are then solved
recursively to obtain the final explicit forms which are used to find the thermodynamic results.  The fact that these
quantities are uncoupled and exactly solvable in the $q=1$ limit facilitates the recursive procedure. In the grandcanonical
ensemble, the final results are computed up to second order in the expansion variable $(1-q)$ whereas in the generalized ensemble
the results are displayed only up to the first order.  Though the results have been calculated and displayed only up to a particular order in the expansion parameter, the procedure can be extended to any arbitrary order in $(1-q)$.

\par

A similar unification procedure is used to create a generalized formulation for the adiabatic ensembles.
The microcanonical ensemble, the isoenthalpic-isobaric ensemble, the adiabatic ensemble with number
fluctuations, and the adiabatic ensemble with number and volume fluctuations belong to this class.  The nonextensive
$q$ generalization of the equipartition theorem and the virial theorem have been obtained in a unified sense.
The ($N$,$P$,$H$) ensemble, ($\mu$,$V$,$\mathsf{L}$) ensemble and the ($\mu$,$P$,$\mathsf{R}$) ensemble have
been studied through the examples of classical nonrelativistic ideal gas and the extreme relativistic ideal gas.
The above mentioned ideal gas models have been solved in the isoenthalpic-isobaric ensemble.  The exact
expression corresponding to the entropy, the enthalpy, the internal energy and the heat capacities at both constant
pressure and at constant volume were found. In the case of ($\mu$,$V$,$\mathsf{L}$) and the ($\mu$,$P$,$\mathsf{R}$)
ensemble an exact solution could not be arrived and so the thermodynamic quantities have been expressed as formal sums.
An interesting observation is made from the study of the ideal gas models in the various ensembles.  The gas models
were exactly solvable in the isothermal-isobaric and the isoenthalpic-isobaric ensembles.  In both these ensembles
we notice that the ratio of the specific heat at constant pressure and the specific heat at constant volume was
found to be independent of the nonextensivity parameter $q$, the number of particles $N$ and the temperature.
This is despite the fact that the individual heat capacities were dependent on these three factors.  Meanwhile
the difference between the heat capacities was found to be dependent on these three factors.

\par

The choice of an ensemble to study a problem in statistical mechanics depends on the physical situation.  The microcanonical
ensemble is used to study completely isolated systems, whereas the canonical ensemble is used to study closed systems which
exchange energy with the surroundings.  The grandcanonical ensemble is used to study systems capable of exchanging particles
with its surroundings.  A system like colloidal particles or macromolecules immersed in a solvent experiences a constant pressure
and so they can be suitably studied using the isothermal-isobaric ensembles.  Depending on whether a fluid under constant pressure
is isothermally or adiabatically confined, we can use the isothermal-isobaric or the isoenthalpic-isobaric ensemble.
The generalized ensemble can be used to analyze systems whose size varies while the pressure is maintained constant.  Typical
examples include molecular clusters in a one component gas and a system of bound ions or molecules on a protein molecule.
Also there are situations in nature where an external electric field, magnetic field or gravitational field acts on a thermodynamic system.  The construction of a class of ensembles comprising of an external field along with the thermal, mechanical and chemical variables, and their consequent application to study problems like rotating black holes and charged polymers like DNA molecules should be worth pursuing.

\par

The unification procedure developed in the current work for the nonextensive $q$-entropy
(\ref{q_entr}) can be extended to other kinds of nonextensive entropies like $\kappa$ entropy
[\cite{GK2001}] and the two parameter entropies [\cite{AS2005},\cite{S2007}]. Such an effort will help us to
understand the change in the structure of thermostatistics due to nonextensivity.  It has been noticed in
Ref. [\cite{JR2}] that there is an inherent pairing between the isothermal and the adiabatic ensembles that is
each isothermal ensemble has an adiabatic counterpart.  These pairs can be identified through the fact that they
have the same heat function.  An interrelation between the isothermal and adiabatic ensemble can be established
through a Laplace transformation.  The canonical ensemble and the microcanonical ensemble have been interrelated in
the nonextensive $q$-statistics  through a $q$-generalization of the Laplace transform [\cite{EL99}]. Currently we are
working on developing a unified formulation of the $q$-Laplace transform to connect any of the isothermal ensemble
to its adiabatic counterpart. The results will appear elsewhere.

%
%
%
\section*{Acknowledgements}
The authors would like to thank Professor Ranabir Chakrabarti
for helpful discussions and motivation, and Dr. Niyaz Ahamed Mandir for providing crucial references. \\
R. Chandrashekar would like to acknowledge the
fellowship received from Council of Scientific and Industrial Research
(India).

%
%
%

\end{document}